\documentclass[10pt,journal,twocolumn,final]{IEEEtran}
\usepackage{amssymb,amsmath,amsthm}
\usepackage{mathrsfs}
\usepackage[utf8]{inputenc}
\usepackage[T1]{fontenc}
\usepackage{graphicx}
\usepackage{epsfig}
\usepackage[english]{babel}
\usepackage{subfigure}
\usepackage{color}
\usepackage{import}
\usepackage{multirow}
\usepackage{cite}
\usepackage{epstopdf}

\usepackage{soul}

\newtheorem{proposition}{Proposition}

\newtheorem{corollary}{Corollary}

\makeatletter
\newcommand*{\rom}[1]{\expandafter\@slowromancap\romannumeral #1@}
\makeatother

\begin{document}

\title{Energy-Recycling Full-Duplex Radios for Next-Generation Networks}

\author{
\authorblockN{Marco Maso,~\IEEEmembership{Member,~IEEE,} Chen-Feng Liu, Chia-Han Lee,~\IEEEmembership{Member,~IEEE,}  \\Tony Q. S. Quek,~\IEEEmembership{Senior Member,~IEEE,} and
Leonardo S. Cardoso,~\IEEEmembership{Member,~IEEE} \\
}

\thanks{This research was supported in part by the A*STAR SERC under Grant 1224104048,  in part by the SUTD-ZJU Research Collaboration under Grant  SUTD-ZJU/RES/01/2014,  in part by the MOE ARF Tier 2 under Grant MOE2014-T2-2-002, and  in part by the Ministry of Science and Technology (MOST), Taiwan under Grant MOST104-2221-E-001-013-MY2.}

\thanks{M. Maso is with the Mathematical and Algorithmic Sciences Lab, Huawei France Research Center, Boulogne-Billancourt 92100, France (e-mail: marco.maso@huawei.com).}

\thanks{C.-F. Liu and C.-H. Lee are with the Research Center for Information Technology Innovation, Academia Sinica, Taipei 115, Taiwan (e-mail: cfliu@citi.sinica.edu.tw; chiahan@citi.sinica.edu.tw).}

\thanks{T. Q. S. Quek is with the Singapore University of Technology and Design, Singapore  487372, and also with the Institute for Infocomm Research, A*STAR, Singapore 138632 (e-mail: tonyquek@sutd.edu.sg).}

\thanks{L. S. Cardoso is with the Université de Lyon, INRIA, INSA-Lyon, CITI-INRIA, F-69621, Villeurbanne, France (e-mail: leonardo.cardoso@insa-lyon.fr).}}

\maketitle
\begin{abstract}
In this work, a novel energy-recycling single-antenna full-duplex (FD) radio is designed, in which a new 3-port element including a power divider and an energy harvester is added between the circulator and the receiver (RX) chain. The presence of this new element brings advantages over the state of the art in terms of both spectral efficiency and energy consumption. In particular, it provides the means of performing both an arbitrary attenuation of the incoming signal, which in turn increases the effectiveness of the state-of-the-art self-interference cancellation strategies subsequently adopted in the RX chain, and the recycling of a non-negligible portion of the energy leaked through the non-ideal circulator. The performance of this architecture is analyzed in a practically relevant 4-node scenario in which 2 nodes operate in FD and 2 nodes in half-duplex (HD). Analytical approximations are derived for both the achievable rates of the transmissions performed by the FD and HD radios and the energy recycled by the FD radios. The accuracy of these derivations is confirmed by numerical simulations. Quantitatively, achievable rate gains up to $40\%$ over the state-of-the-art alternatives, in the considered scenario, are highlighted. Furthermore, up to $50\%$ of the leaked energy at the circulator, i.e., $5\%$ of the energy of the transmitted signal, can be recycled.
\end{abstract}


\section{Introduction} \label{Sec: Introduction}

\IEEEPARstart{S}{ince} the introduction of personal mobile services, wireless traffic demands have increased continuously. This increase has picked up pace in the last decade due to the popularity of mobile applications and the market penetration of mobile devices with networking capabilities. In a moment when wireless research has continuously closed the gap between the achievable and theoretical capacity bounds, innovative solutions are needed to further enhance the performance of current networks. In this context, one very appealing approach that has regained traction lately is the in-band full-duplex (FD)~\cite{art:zhang15}. Devices adopting this approach can simultaneously transmit and receive signals in the same frequency band. 

FD radios can theoretically achieve a two-fold throughput improvement without extra antennas or band over their half-duplex (HD) counterparts. Among the potential solutions to implement FD radios for future networks, the single-antenna architectures have recently gained momentum~\cite{Hon:2014:5GFD}. This happens for several reasons. These systems take significantly less physical space on the devices than their multi-antenna counterparts, being less demanding in terms of the form factor. This is of utmost importance when the possible size of the FD device is limited either by physical constraints, e.g., a sensor, or by practical and commercial purposes, e.g., a portable device. Remarkably, these architectures have been shown to be able to achieve the claimed theoretical performance both in terms of academic prototypes and commercially available products~\cite{onl:kumu}. The further element in favor of single-antenna FD architectures is that they share the same radio-frequency (RF) circuitry basis with HD systems. Therefore, very little hardware changes are needed to implement a single-antenna FD radio as compared to a HD one.

The potential performance enhancement brought by the FD approach depends on how much self-interference (SI) can be subtracted from the received signal~\cite{art:sabharwal14}. This aspect is especially relevant in the context of single-antenna FD radios, where the self-interference cancellation (SIC) capability of the device is significantly hindered by practical hardware limitations, e.g., the non-ideality of the antenna circulator~\cite{conf:bharadia13, conf:knox12}. From a qualitative point of view, the SI in single-antenna FD radios has two main components: 1) signal leakage between the transmit (TX) and receive (RX) ports of the circulator, and 2) signal received by the antenna after a propagation in a multi-path environment. In this regard, it is worth noting that the signal leakage from the circulator not only generates SI but also reduces the energy efficiency of the FD radio since part of the power invested by the latter for the transmission is wasted due to the leakage.

\subsection{Related Works}
The design of SIC algorithms has been widely addressed before. Notable examples in the literature are the solutions based on either analog~\cite{conf:knox12, conf:Jain, conf:bharadia13, Chen:1998:DFD, conf:Phun} or digital signal processing~\cite{conf:li12, art:ahmed14, art:korpi14, conf:korpi14_2} SIC, and the so-called spatial SIC~\cite{conf:Choi:10,art:Ngo:14,art:Suraweera:14}. In general, performing part of the SIC in the analog domain reduces the problems of the saturation of the RX amplifiers and the low dynamic range at the analog-to-digital converter (ADC)~\cite{conf:bharadia13}. Several approaches have been proposed in this context. A solution based on a balanced feed network in which the transmitted signal is fed to the TX antenna via two paths such that their corresponding SI signals are 180 degrees out of phase is proposed in~\cite{conf:knox12}. Alternative strategies are proposed in~\cite{Chen:1998:DFD} and~\cite{conf:Phun}, where the adoption of a RF echo canceler and a phase shifter to tune a reference signal in order to match it with the SI signal are proposed, respectively. The best SIC capabilities are certainly offered by hybrid solutions adopting both analog and digital signal processing~\cite{conf:Jain, conf:bharadia13}.
 A quantitatively remarkable result in this sense is achieved in~\cite{conf:bharadia13}, where a novel single-antenna FD radio architecture able to achieve 110\,dB of SIC for a transmission of an orthogonal frequency-division multiplexing (OFDM) signal over a bandwidth of 80~MHz is proposed. This solution, essentially based on  a hybrid analog and digital cancellation algorithm, provides a very effective way of implementing a FD transceiver without significantly increasing both its size and cost. The market potential of such single-antenna FD radio is certainly non-negligible. Motivated by these achievements, prototypes and market-ready products have already been showcased and proposed to demonstrate the feasibility of single-antenna FD transmissions in real-life scenarioss~\cite{onl:kumu, Bharadia:2014:RFD}, confirming that real-time FD radios can effectively operate in different environmental conditions.

\subsection{Our Contribution}

The effectiveness of state-of-the-art SIC algorithms are subject to constraints on the TX power of the FD radio, and hence, the residual SI~\cite{conf:bharadia13, conf:knox12}. As a matter of fact, the intensity of the SI can be brought down to the noise floor only if the TX power is below a certain threshold, which is determined by the nature of the SIC algorithm. Conversely, residual SI appears in the RX chain and the signal-to-interference-plus-noise ratio (SINR) of the incoming received signal decreases. This can significantly degrade the throughput of the FD radio. In this work, we approach this problem, considering that the range of TX powers of the FD radio may span values above the threshold for complete SIC. To deal with this issue, a novel FD radio architecture is designed and proposed in this work. The core novelty of this architecture, as compared to the state-of-the-art alternatives, is the introduction of a one-way 3-port composite element, composed of a variable gain power divider and an RF energy harvester (EH),  between the circulator and the RX chain. In practice, the 3-port element splits the signal coming from the circulator into two parts, i.e., an information component (IC) to be decoded by the RX chain and an energy component (EC) to be harvested. This has a two-fold advantage on the performance of the FD device. Firstly, the intensity of the SI observed in the RX chain can be arbitrarily reduced, regardless of the TX power of the FD radio, in turn both restoring the effectiveness of the state-of-the-art SIC algorithms and increasing the spectral efficiency of the incoming link. Secondly, the energy efficiency of the FD radio can be increased by recycling a portion of the energy leaked at the circulator, otherwise wasted.  However, this performance gain may come at a price. In fact, on one hand the combined effect of the 3-port element and SIC provides an effective SINR enhancement for the useful signal carried by the IC. On the other hand,  the power of the latter may significantly decrease as compared to the noise floor during these operation, inducing an intrinsic signal-to-noise ratio (SNR) reduction. In practice, a trade-off, a function of the splitting ratio adopted by the power divider, exists. As a consequence, a careful performance analysis needs to be performed to optimize the relevant parameters of the proposed architecture and assess its merit as compared to the state of the art.

A practically relevant case study is identified to achieve this goal. Accordingly, we consider a hybrid FD/HD four-node setting in which two FD nodes play the role of server nodes (SN) while the remaining two play the role of attached nodes (AN) operating in HD. In this context, we divide the communication between the SNs and the ANs into two directions: 1) forward (or ``downlink''), and 2) backward (or ``uplink''). Additionally, we assume that the two SNs exchange information, regardless of the direction of the communication. The rationale behind this setting is that it perfectly captures the main elements of what is currently considered the best candidate for future FD-enabled networks, i.e., a system in which FD and HD devices operate side by side~\cite{art:goyal15}.  Interestingly, this rather general setting can model several different real-life scenarios. Possible relevant examples are: (i) a scenario in which the proposed architecture is used at two mobile devices engaging in a device-to-device (D2D) communication, while being served by two base stations (BSs), or (ii) a scenario in which the proposed architecture is used at two BSs exchanging signaling via a wireless backhaul while serving two mobile devices. In this regard, we would like to note that both D2D communications and wireless backhauling are indeed two of the envisioned applications for the FD technology both in academic~\cite{art:sabharwal14, art:li2015} and industrial~\cite{onl:kumu} contexts. Furthermore, we assume that the SNs adopt orthogonal frequency-division multiple access (OFDMA) to serve their ANs in the same time slot and frequency band. This choice has a two-fold motivation. Firstly, this waveform, together with its many variants (e.g., filtered OFDM~\cite{onl:huawei}), is one of the most likely candidates for the air interface design of future 5G networks for its potential in terms of high data rates, mobility management, transmission diversity exploitation potential, spectral efficiency, robustness against inter-symbol interference, and simple receiver architecture requirements~\cite{art:Hwang:09}. Secondly, many state-of-the-art works on FD radios are actually based on OFDM, e.g.,~\cite{conf:bharadia13}, due to the aforementioned reasons. As a consequence, considering the same waveform as these works gives us the possibility to compare our results against the state of the art in a very direct and fair way. Finally, in order to frame a scenario in which the potential of the hybrid FD/HD is fully exploited, we impose that the FD communications between the two SNs do not affect the HD communications between SNs and ANs. In other words, we explicitly account for an intrinsic inter-node interference reduction at the PHY layer, similarly to what is advocated in contributions such as~\cite{art:bai13, DBLP:journals/corr/DuTDS15}.  In practice, we will impose that the signaling between the two FD SNs is exchanged an overlay technique specifically tailored for OFDM-based communications, the so-called cognitive interference alignment (CIA)~\cite{art:Maso13:VT}.

Subsequent to the definition of the reference setting, we analytically derive the exact/approximated achievable rate of each link in the considered system. Additionally, we specifically analyze the aforementioned trade-off between SINR enhancement and SNR decrease by studying how the splitting ratio of the power divider impacts the achievable rates. In this context, we compute closed-form expressions for the sets of values of the splitting ratio for which the proposed architecture outperforms the state-of-the-art alternatives. In particular, we provide the optimal value of such parameter as a function of the TX power of the device. Finally, we validate our derivations by means of numerical simulations and show that for the considered reference setting: 1) achievable rate gains of up to $40\%$ over the state-of-the-art alternatives can be obtained, and 2) up to $50\%$ of the leaked energy at the circulator, i.e., $5\%$ of the energy of the transmitted signal, can be recycled. 

The remainder of the paper is organized as follows.  The novel FD radio architecture is introduced in Sec.~\ref{Sec: FD architecture}. The system setup is described in Sec.~\ref{Sec: System model}. The forward and backward transmission modes are analyzed in Sec.~\ref{Sec: Downlink} and Sec.~\ref{Sec: Uplink}, respectively. Finally, numerical results are presented in Sec.~\ref{Sec: Numerical results} and conclusion in Sec.~\ref{Sec: Conclusions}. 

Throughout this work, the mathematical notation adopted is as follows. We denote matrices as boldface upper-case letters, vectors as boldface lower-case letters, and we let $(\cdot)^\mathrm{H}$ be the conjugate transpose of a vector/matrix. All vectors are columns, unless otherwise stated. In particular, $\mathbf{I}_N$ is the $N\times N$ identity matrix, and $\mathbf{0}_{L\times N}$ is the  $L\times N$ zero matrix. We denote $\left[\mathbf{H}\right]_{mn}$ and $\left[\mathbf{h}\right]_{n}$ as the element in the $m$th row and the $n$th column of a matrix and the $n$th element of a vector, respectively. Finally, $\mbox{diag}(\mathbf{h})$ is a diagonal matrix constructed from the vector $\mathbf{h}$.
\section{Energy-recycling FD Architecture}\label{Sec: FD architecture}

\begin{figure}[!h]
	\centering
	{\def\svgwidth{0.83\columnwidth}
		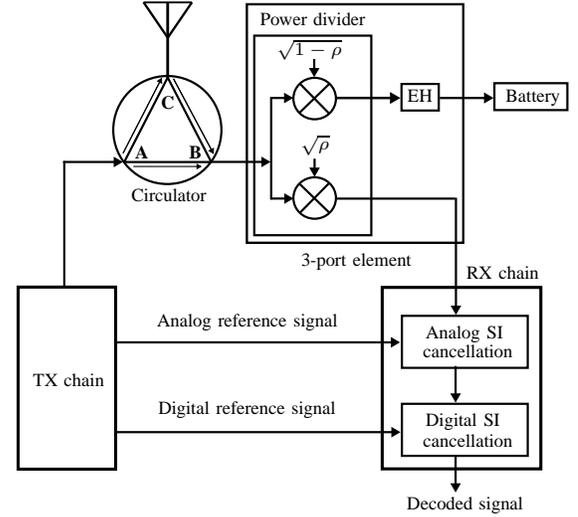}
	\caption{Novel FD architecture.}
	\label{Fig: FD archi} 
\end{figure}

Let us now focus on the architecture of the FD SN. We start by recalling that state-of-the-art SIC algorithms for FD radios are limited in the amount of SI they can cancel. In practice, the full effectiveness of the adopted SIC algorithms, which aim at reducing the intensity of the SI at least at the same level of the noise floor, is guaranteed only if the TX power is below a certain threshold~\cite{conf:bharadia13}. We denote this upper bound as $P_{\mathrm{th}}$. Now, let us assume that the FD device is able to induce an arbitrary attenuation of the SI such that the power of the latter is lower than $\alpha_{\mathrm{c}}P_{\mathrm{th}}$, with $\alpha_{\mathrm{c}}$ defined as the ratio between the power of the signal leakage and the TX power. In this case, the SIC technique would provide its best cancellation performance. As a consequence, significant SINR gains w.r.t.~state-of-the-art solutions could be achieved, thanks to the joint effect of the arbitrary attenuation and the SIC. The design of the novel FD architecture proposed in this work starts from these considerations. Consider the FD radio architecture depicted in Fig.~\ref{Fig: FD archi}. 
As seen in Fig.~\ref{Fig: FD archi}, a 3-port composite element is introduced in the new architecture between the port \textbf{B} of the circulator and the RX chain, and hybrid SIC approach as in~\cite{conf:bharadia13} is assumed. Interestingly, this element can be built using off-the-shelf components easily available on the market. In particular, it is composed of:
\begin{itemize}
	\item{\textbf{Power Divider:}} The input signal to the 3-port element is first fed to an adjustable gain power divider and split into two with the $\sqrt{\rho}$ and $\sqrt{1-\rho}$ power ratios. We refer to $\rho$ as the power splitting ratio. This component is commonly considered at the core of many wireless power transfer applications, as the enabler of the so-called power splitting approach~\cite{art:zhou13arch}. In particular, it can be implemented by means of Wilkinson power dividers and balanced-unbalanced (balun) transformers, commonly developed and studied components in the RF circuitry domain. 
In this context, variable gain variants of such power dividers/balun have been the subject of intense research in last years. Relevant examples of the outcome of these efforts are the variable gain Wilkinson power divider proposed and implemented in microstrip for microwave applications in~\cite{art:wu2010} or, more recently, the variable power division balun proposed and implemented in microstrip in~\cite{art:zhang2014}.  In the proposed architecture, the outputs of the divider are two attenuated replicas of the input signal, i.e., the IC and the EC. After the power division, the IC is fed to the RX chain connected to the second port of the 3-port element, whereas the EC is fed to the second component of the 3-port element, i.e., the EH circuitry. In this regard, we note that no power division occurs whenever $\rho=1$, i.e., the entirety of the input signal to the power divider is fed to the RX chain of the FD radio. This corresponds to a situation in which the proposed FD radio operates exactly as the state-of-the-art solution~\cite{conf:bharadia13}, as discussed in the following.
	\item{\textbf{Energy Harvester:}} The EH circuitry converts a time varying signal, i.e., the EC, into a direct current (DC) signal suitable for battery recharging or powering circuits. 	The structure and operations performed by a state-of-the-art EH~\cite{art:Visser12, GabrielAbadal2014-02-12} are:
	\begin{enumerate}
		\item The input signal is rectified by means of a device commonly referred to as \textit{rectifier}. In general, this operation is performed by means of suitable diodes. In particular, p-n junction diodes are adopted when the signal has frequencies in the kHz-MHz range, whereas devices with shorter transit times and lower intrinsic capacitances, such as the gallium arsenide (GaAs) Schottky diodes, are adopted when the signal has frequencies in the GHz-THz range~\cite{GabrielAbadal2014-02-12};
		\item The rectified signal is filtered by means of a second-order low-pass filter to obtain a DC voltage;
		\item Finally, a DC-to-DC converter, e.g., an \emph{unregulated buck-boost converter} operating in the discontinuous conduction mode~\cite{art:Visser12}, is usually adopted to adapt the rectified voltage to the level required by the application load, e.g., a storage device, connected to the third port of the 3-port element in Fig.~\ref{Fig: FD archi}.
	\end{enumerate} 
	The obtained voltage by means of this procedure can charge a battery within a range of few Volts. The efficiency of the overall RF-to-DC conversion provided by the EH can be modeled by a factor $\beta\in [0,1]$, obtained as the ratio of the DC output energy over the RF input energy~\cite{art:Visser12, art:zhang13}. It is worth noting that the state-of-the-art RF EH is already capable of delivering remarkable conversion efficiencies, i.e., $\eta\geq 50\%$, and is ready for commercial usage~\cite{art:Visser12, art:Xie:13, prod:P2110}.  Naturally, the conversion efficiency strongly depends on how much the EH can be tailored to the specific architecture/application. In this context, an appropriate tuning of the EH corresponds to the adoption of a suitable diode able to operate with signals at the frequencies of interest. As a consequence, every RF EH can be appropriately tuned as long as its components are suitably chosen~\cite{GabrielAbadal2014-02-12}.
\end{itemize}
The proposed architecture brings two main advantages over the state of the art: first, it allows to increase the upper bound on the TX power for the FD radio while guaranteeing the effectiveness of the preexisting SIC algorithms; second, it recycles a portion of the wasted energy at the circulator. It should be noted that such architecture also offers an implicit additional advantage in the form of general RF energy harvesting capabilities. In particular, and similar to what is done in classic wireless power transfer settings~\cite{art:zhang13, art:xie12, art:liu13}, RF energy could always be harvested from the environment by the proposed FD architecture by setting $\rho=0$, whenever no information signal is received. Concerning the latter aspect, we would like to note that differently from other FD architectures that resort to RF energy harvesting to increase their energy efficiency~\cite{art:Zeng:15}, no restrictions on the usage that may be done of the recycled energy are enforced in this contribution. Indeed, several reasonable strategies to make use of such energy could be envisioned for the proposed architecture, e.g., the one proposed in~\cite{art:Zeng:15}.  However any choice in this sense would be arbitrary and out of the scope of this work.  Thus, the identification of possible strategies for the proposed FD architecture to use the recycled energy is deferred to a further contribution.

\subsection{Impact on the Performance of the FD Radio}

The operating scenario in which the proposed architecture unveils its potential is when the TX power of the FD radio, denoted by $P$, is larger than $P_{\mathrm{th}}$. In this case, the residual SI suffered by the state-of-the-art solution after the SIC would have power $\frac{\rho P N_0}{P_{\mathrm{th}}}$. Hence, the choice of an appropriate value of $\rho$, i.e., $\rho\leq\frac{P_{\mathrm{th}}}{P}$, could reduce the power of the signal coming from the circulator such that the full effectiveness of the subsequent SIC algorithms is restored. Conversely, no advantage would be brought by the proposed FD architecture over the state of the art when $P \leq P_{\mathrm{th}}$, i.e., $\rho=1$ is optimal. Naturally, as previously discussed, the power divider targets the entirety of its input signal. This results in a non-negligible SNR reduction w.r.t.~the state-of-the-art solution, potentially affecting the performance of the device. To provide the best operating point, the choice of an optimal $\rho$ must be made accounting for both gains and losses associated to SINR enhancement and SNR reduction. This aspect will be further discussed in the following. 

Finally, it is worth noting that the performance of the FD radio is constrained not only by the limitation of the state-of-the-art SIC solutions but also by practical hardware limitations that affect the RX chain of every radio device. In practice, the circuitry performing the digital signal processing in the RX chain can easily saturate if the power of its input signal is too large. Let $P_{\mathrm{sat}}\geq P_{\mathrm{th}}$ be the TX power of the FD device for which such saturation occurs. Thus, in general, the decoding of the IC can be performed by the FD radio only when $\rho\leq\frac{P_{\mathrm{sat}}}{P}$.

\section{System Setup}\label{Sec: System model}

\begin{figure}[!h]
	\centering
	\subfigure[Forward phase.]
	{\def\svgwidth{.3\columnwidth}
		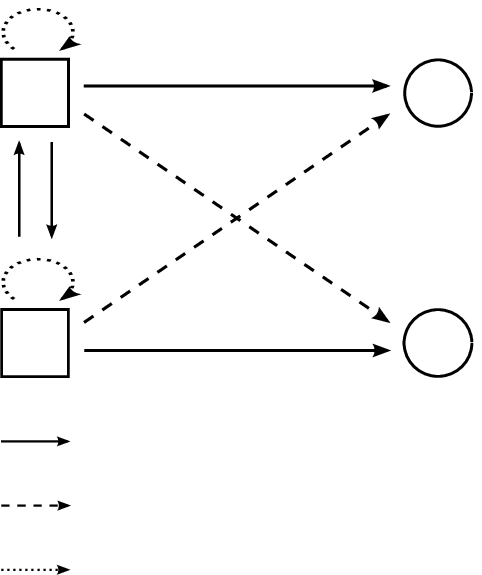\label{Fig: P1 signal flow} }
\hspace{1cm}	
	\subfigure[Backward phase.]
	{\def\svgwidth{.3\columnwidth}
		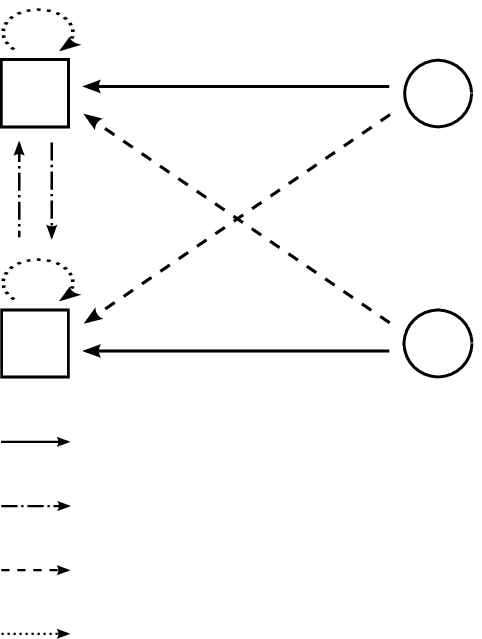
		\label{Fig: P2 signal flow}}
	\caption{Communication phases between the SNs and the ANs.}
	\label{Fig: System model}
\end{figure}

Consider the hybrid FD/HD four-node scenario described in Sec.~\ref{Sec: Introduction}, in which two FD SNs serve two HD ANs, while sharing information/signaling over a wireless in-band link. In particular, and without loss of generality, we assume that the communication between the FD SNs and the HD ANs occurs in the time-division duplexing (TDD) mode. The two phases of this communication are illustrated in Fig.~\ref{Fig: System model}. 
%
%
During the forward phase, the $i$th SN simultaneously communicates with both the $i$th AN and the $j$th SN. In the meanwhile, the $i$th AN only receives the signals from the two SNs, without transmitting. Conversely, in the backward phase, the $i$th AN communicates with its corresponding SN, without receiving information from the latter. At the same time, the $i$th SN communicates with the $j$th SN while receiving signals from all the other nodes. 
We assume frequency-selective block-fading channels with $l+1$ taps between all the devices, distributed as independent complex Gaussian random variables with zero mean and variance given by $\xi_{n}^{2}\in\mathbb{R}$, depending on the power delay profile (PDP) of the considered channel. We denote the vectors representing the channel between the $i$th-SN-to-$j$th-AN link, the channel between the SNs, and the multi-path channel experienced by the SI signal at the $j$th SN as $\mathbf{h}_{ij}=\left[h_{ij}(0),\cdots,h_{ij}(l)\right]^{\mathrm{T}}\in\mathbb{C}^{l+1}$, $\mathbf{h}_{s}=\left[h_{s}(0),\cdots,h_{s}(l)\right]^{\mathrm{T}}\in\mathbb{C}^{l+1}$, and  $\mathbf{h}^m_{j}=\left[h_{j}^{m}(0),\cdots,h^{m}_{j}(l)\right]^{\mathrm{T}}\in\mathbb{C}^{l+1},\forall\,i,j\in\left\{1,2\right\}$, respectively, with $h_{ij}(n)$, $h_{s}(n)$, and $h_{j}^{m}(n)\in\mathcal{CN}(0,\xi_n^2),\,\forall\,n\in[0,l]$. Assuming no RF impairments and perfect synchronization at the receiver, the channel matrix which models the convolution of the signal transmitted by the $i$th SN and $\mathbf{h}_{ij}$ can be written as
\begin{equation}
\mathbf{H}_{ij}=\left[
\begin{array}{cccccc}
h_{ij}(0)&0&\cdots&h_{ij}(l)&\cdots&h_{ij}(1)
\\h_{ij}(1)&\ddots&\ddots&\ddots&\ddots&\vdots
\\\vdots&\ddots&\ddots&\ddots&\ddots&h_{ij}(l)
\\h_{ij}(l)&\ddots&\ddots&h_{ij}(0)&\ddots&0
\\\vdots&\ddots&\ddots&\ddots&\ddots&0
\\0&\cdots&h_{ij}(l)&\cdots&\cdots&h_{ij}(0)
\end{array}
\right].
\end{equation}
Analogously, $\mathbf{H}_{s}$ and $\mathbf{H}_{j}^{m}$ are constructed from $\mathbf{h}_{s}$ and $\mathbf{h}_{j}^{m}$, respectively. At this stage, we assume that perfect channel state information (CSI) is available at each device. As discussed in Sec.~\ref{Sec: Introduction}, we assume that the communication between SNs and ANs is based on OFDMA with $N$ sub-carriers and $L$ cyclic prefix (CP) symbols. This assumption guarantees three conditions: 1) the frequency-selectivity of the channels can be effectively dealt with at the receivers with simple equalization procedures, 2) the considered waveform is consistent with the trends envisioned for the 5G air interface design~\cite{onl:huawei}, and 3) the resulting scenario is compliant with what is considered in the reference state-of-the-art case~\cite{conf:bharadia13}. We denote $\mathscr{N}_{i}$ as the set of sub-carriers allocated to the $i$th SN-to-AN link, where $\mathscr{N}_1\cup\mathscr{N}_2=\left\{1,\cdots,N\right\}$ and $\mathscr{N}_1\cap\mathscr{N}_2=\emptyset$. In this context, the signaling is performed by means of a block transmission with block size $N+L$ symbols, as detailed in the following. 

\section{The Forward Phase}\label{Sec: Downlink} 
\subsection{Signal Model}

According to the sub-carrier allocation for the OFDMA transmission, the information transmitted by the $i$th SN in the $t$th OFDM block, i.e., $\mathbf{u}_{i,\mathrm{\mathrm{o}}}[t]\in\mathbb{C}^{N}$, can be expressed as
\begin{equation}
\big[\mathbf{u}_{i,\mathrm{o}}[t]\big]_n=\begin{cases}
\big[\mathbf{u}_{i,\mathrm{o}}[t]\big]_n\in\mathbb{C},&\hspace{.5mm}n\in \mathscr{N}_i,\\
0, &\mathrm{otherwise},
\end{cases}
\label{Eq:Down_OFDM_Information}
\end{equation}
with covariance matrix $\mathbf{P}^{{\rm F}}_{i,\mathrm{o}}\in\mathbb{C}^{N\times N}$. Let $\mathbf{F}$ be the normalized $N$-point discrete Fourier transform (DFT) matrix, i.e., $\left[\mathbf{F}\right]_{mn}=\frac{1}{\sqrt{N}}\exp\left({-\frac{j2\pi (m-1)(n-1)}{N}}\right),~\forall\,m,n\in\{1,\cdots,N\}$, and
\begin{equation*}
\mathbf{A}=\left[
\begin{array}{rc}
\mathbf{0}_{L\times (N-L)}&\mathbf{I}_{L}
\\\mathbf{I}_{N}&
\end{array}
\right]
\end{equation*}
a CP insertion matrix, then the signal transmitted by the $i$th SN to the $i$th AN in the forward phase is obtained as $\mathbf{x}_{i,\mathrm{o}}[t]=\mathbf{A}\mathbf{F}^{-1}\mathbf{u}_{i,\mathrm{o}}[t] \in \mathbb{C}^{N+L}$, such that $\mbox{Tr}\big(\mathbb{E}[\mathbf{x}_{i,\mathrm{o}}[t]\big(\mathbf{x}_{i,\mathrm{o}}[t]\big)^{\mathrm{H}}]\big)=(N+L)P_{\mathrm{o}}$. Let $\mathbf{x}_{i,\mathrm{b}}[t]$ be the signaling transmitted from the $i$th to the $j$th SN in the $t$th block, detailed in the following for simplicity. Thus, we can write the overall transmitted signal by the $i$th SN in the $t$th block as $\mathbf{x}_{i}[t]=\mathbf{x}_{i,\mathrm{o}}[t]+\mathbf{x}_{i,\mathrm{b}}[t]$. Now, if we define $\alpha_{ij}$ as the path loss attenuation between the $i$th SN and the $j$th AN, then the $t$th block is received by the $i$th AN as
\begin{align}
\mathbf{y}_{i,\mathrm{o}}^{{\rm F}}[t]&=\sqrt{\alpha_{ii}}\left(\mathbf{H}_{ii}^{\nabla}\mathbf{x}_{i}[t]+\mathbf{H}^{\triangle}_{ii}\mathbf{x}_{i}[t-1]\right)\notag
\\&\quad+\sqrt{\alpha_{ji}}\left(\mathbf{H}_{ji}^{\nabla}\mathbf{x}_{j}[t]+\mathbf{H}^{\triangle}_{ji}\mathbf{x}_{j}[t-1]\right)
+\mathbf{w}_{i,\mathrm{o}}[t],
\label{Eq:Down_Received_Signal_at_AN}
\end{align}
with $\mathbf{w}_{i,\mathrm{o}}[t]\in\mathcal{CN}(\mathbf{0},N_{0}\mathbf{I}_{N+L})$ additive white Gaussian noise (AWGN), and where $\mathbf{H}_{ij}^{\nabla}$ and $\mathbf{H}_{ij}^{\triangle}$ are the lower and the upper triangular part of $\mathbf{H}_{ij}$ responsible for the inter-symbol and inter-block interference (ISI and IBI), respectively. In this regard, we note that all the received signals by the FD radio at time $t$ appear in \eqref{Eq:Down_Received_Signal_at_AN} for the sake of completeness, i.e., the signals transmitted by the serving and the interfering SN, including both the $t$th OFDM block and the IBI component resulting from the multi-path propagation of the $(t-1)$th OFDM block, and received with the $t$h block. In order to decode the received signal, the $i$th AN performs a legacy OFDMA demodulation and get
\begin{align}
\mathbf{y}_{i,\mathrm{o}}^{{\rm F}}[t]&=\mathbf{D}_{i}\mathbf{F}\mathbf{B}\Big(\sqrt{\alpha_{ii}}\left(\mathbf{H}_{ii}^{\nabla}\mathbf{x}_{i}[t]+\mathbf{H}^{\triangle}_{ii}\mathbf{x}_{i}[t-1]\right)\notag
\\&+\sqrt{\alpha_{ji}}\left(\mathbf{H}_{ji}^{\nabla}\mathbf{x}_{j}[t]+\mathbf{H}^{\triangle}_{ji}\mathbf{x}_{j}[t-1]\right)
+\mathbf{w}_{i,\mathrm{o}}[t]\Big),
\label{Eq:Down_OFDM_Demodulated_Signal_at_AN}
\end{align}
where $\mathbf{B}=[\mathbf{0}_{N\times L} \hspace{2mm} \mathbf{I}_{N}]$ is the CP removal matrix, and $\mathbf{D}_{i}$ is the $N\times N$ sub-carrier selection matrix for $\mathscr{N}_{i}$, i.e.,
\begin{equation*}
[\mathbf{D}_{i}]_{mn}=\begin{cases}
1,&\hspace{.5mm}m=n\in\mathscr{N}_{i},\\
0, &\mathrm{otherwise}.
\end{cases}
\end{equation*}
In practice, the signal obtained after this process can be written as
\begin{align}
\mathbf{y}_{i,\mathrm{o}}^{{\rm F}}[t]&=\sqrt{\alpha_{ii}}\mathbf{D}_{i}\mathbf{F}\mathbf{B}\mathbf{H}_{ii}^{\nabla}\left(\mathbf{A}\mathbf{F}^{-1}\mathbf{u}_{i,\mathrm{o}}[t]+\mathbf{x}_{i,\mathrm{b}}[t]\right)\notag
\\&\quad+\sqrt{\alpha_{ji}}\mathbf{D}_{i}\mathbf{F}\mathbf{B}\mathbf{H}_{ji}^{\nabla}\mathbf{x}_{j,\mathrm{b}}[t]
+\mathbf{D}_{i}\mathbf{F}\mathbf{B}\mathbf{w}_{i,\mathrm{o}}[t].\label{Eq:Down_OFDM_Demodulated_WithoutIBI_at_AN}
\end{align}
Switching our focus to the signaling between the two SNs, we note that it should not generate undesired interference during the OFDM decoding of $\mathbf{u}_{i,\mathrm{o}}[t]$ at the $i$th AN, as discussed in Sec.~\ref{Sec: Introduction}. Interestingly, the structure of the OFDMA signal inherently offers a means to achieve this goal~\cite{art:Maso13:VT}. As a matter of fact, $\mathbf{x}_{i,\mathrm{b}}[t]$ does not interfere with $\mathbf{u}_{i,\mathrm{o}}[t]$ at the $i$th AN only if
\begin{align}
\begin{cases}
\mathbf{D}_{i}\mathbf{F}\mathbf{B}\mathbf{H}_{ii}^{\nabla}\mathbf{x}_{i,\mathrm{b}}[t]=\mathbf{0},
\\\mathbf{D}_{j}\mathbf{F}\mathbf{B}\mathbf{H}_{ij}^{\nabla}\mathbf{x}_{i,\mathrm{b}}[t]=\mathbf{0},
\end{cases}\forall\,i,j\in\left\{1,2\right\}\mbox{ and }i\neq j.\label{Eq:Down_Backhaul_Constraints}
\end{align}
In practice, \eqref{Eq:Down_Backhaul_Constraints} is satisfied if $\mathbf{x}_{i,\mathrm{b}}[t]\in\mbox{null}\big(\mathbf{D}_{i}\mathbf{F}\mathbf{B}\mathbf{H}_{ii}^{\nabla}+\mathbf{D}_{j}\mathbf{F}\mathbf{B}\mathbf{H}_{ij}^{\nabla}\big)$ with $\mbox{dim}\big(\mbox{null}\big(\mathbf{D}_{i}\mathbf{F}\mathbf{B}\mathbf{H}_{ii}^{\nabla}+\mathbf{D}_{j}\mathbf{F}\mathbf{B}\mathbf{H}_{ij}^{\nabla}\big)\big)=L$. We know from~\cite{art:Maso13:VT} that \eqref{Eq:Down_Backhaul_Constraints} can always be satisfied if perfect CSI is available at the transmitter. This means that the $i$th SN can transmit up to $L$ independent information streams by precoding them with a matrix whose column space is equal to $\mbox{null}\big(\mathbf{D}_{i}\mathbf{F}\mathbf{B}\mathbf{H}_{ii}^{\nabla}+\mathbf{D}_{j}\mathbf{F}\mathbf{B}\mathbf{H}_{ij}^{\nabla}\big)$, i.e., $\mathbf{u}_{i,\mathrm{b}}\left[t\right] \in \mathbb{C}^{L}$ with covariance matrix $\mathbf{P}^{\rm F}_{i,\mathrm{b}}\in\mathbb{C}^{L\times L}$. Now, given an arbitrary semi-unitary matrix $\mathbf{\Gamma}_{i}^{{\rm F}}\in\mathbb{C}^{(N+L)\times L}$ that satisfies 
\begin{align}
\begin{cases}
\mathbf{D}_{i}\mathbf{F}\mathbf{B}\mathbf{H}_{ii}^{\nabla}\mathbf{\Gamma}_{i}^{{\rm F}}=\mathbf{0},\\
\mathbf{D}_{j}\mathbf{F}\mathbf{B}\mathbf{H}_{ij}^{\nabla}\mathbf{\Gamma}_{i}^{{\rm F}}=\mathbf{0},
\end{cases}\label{Eq:Down_Backhaul_Constraints_Relaxed} 
\end{align}
obtainable by means of matrix operations such as the LQ factorization, we can express the signaling of the $i$th SN as
\begin{align}
\mathbf{x}_{i,\mathrm{b}}\left[t\right]=\mathbf{\Gamma}_{i}^{{\rm F}}\mathbf{C}_{i}^{{\rm F}}\mathbf{u}_{i,\mathrm{b}}\left[t\right] \in \mathbb{C}^{N+L},
\label{Eq:Down_Backhaul_Transmitted_Signal}
\end{align}
such that $\mbox{Tr}\big(\mathbb{E}[\mathbf{x}_{i,\mathrm{b}}[t]\big(\mathbf{x}_{i,\mathrm{b}}[t]\big)^{\mathrm{H}}]\big)=(N+L)P_{\mathrm{b}}$ and with $\mathbf{C}_{i}^{{\rm F}}\in\mathbb{C}^{L\times L}$ as a matrix offering auxiliary degrees of freedom to the $i$th SN, to be designed according to criteria of interest as detailed in the following. Additionally, we note that $P_{\mathrm{b}}$ is chosen such that the TX power budget per symbol at the SN, i.e., $P$, is not exceeded, that is $P_{\mathrm{b}}=P-P_{\mathrm{o}}$. 

\subsection{Performance of the Transmission in the Forward Phase}\label{Subsec: Down_OFDM}

We first consider the rates achieved by the transmission in the forward phase. Given that the signaling between the SNs satisfy \eqref{Eq:Down_Backhaul_Constraints}, we can rewrite \eqref{Eq:Down_OFDM_Demodulated_WithoutIBI_at_AN} as
\begin{align}
\mathbf{y}_{i,\mathrm{o}}^{{\rm F}}[t]&=
\sqrt{\alpha_{ii}}\mathbf{D}_{i}\mbox{diag}\big(\mathbf{\tilde{h}}_{ii}\big)\mathbf{u}_{i,\mathrm{o}}[t]+\mathbf{D}_{i}\mathbf{F}\mathbf{B}\mathbf{w}_{i,\mathrm{o}}[t],\label{Eq:Down_OFDM_Demodulated_WithoutInterference_at_AN}
\end{align}
where $\mathbf{\tilde{h}}_{ii}=\sqrt{N}\mathbf{F}\left[\mathbf{h}_{ii}^{\mathrm{T}},\mathbf{0}_{1\times(N-l-1) }\right]^{\mathrm{T}}.$
Like the conventional OFDMA scheme, the $i$th SN transmits $\lvert\mathscr{N}_{i}\rvert$ independent information streams to the $i$th AN via $\lvert\mathscr{N}_{i}\rvert$ parallel Gaussian channels. Thus, the achievable rate over the link between the $i$th SN to the $i$th AN in the forward phase is given by 
\begin{align}
&\hspace{-0.5em}R_{i,\mathrm{o}}^{{\rm F}}=\frac{1}{N+L}\sum_{n\in\mathscr{N}_{i}}\log_2\left(1+\frac{\alpha_{ii}[\mathbf{P}_{i,\mathrm{o}}^{{\rm F}}]_{nn}\lvert[\mathbf{\tilde{h}}_{ii}]_{n}\rvert^2}{N_{0}}\right),\label{Eq:Down_OFDM_Rate}
\end{align}
where
%
%
$\big[\mathbf{P}_{i,\mathrm{o}}^{{\rm F}}\big]_{nn}=\left(\kappa_{1,i}-\frac{N_{0}}{\alpha_{ii}\lvert[\mathbf{\tilde{h}}_{ii}]_{n}\rvert^2}\right)^{+},~\forall\,n\in\mathscr{N}_{i},$
%
%
is the water-filling (WF) power allocation solution, with $\kappa_{1,i}$ chosen such that $\mbox{Tr}\big(\mathbf{P}_{i,\mathrm{o}}^{{\rm F}}\big)=NP_{\mathrm{o}}$~\cite{TseWireless}.
%
%

\subsection{Performance of the Signaling between SNs}\label{Subsec: Down_Backhaul}
We now focus on the achievable rate over the link between the two SNs in the considered scenario. The received signal at the $j$th SN, including both the SI and the signal from the other SN, in the $t$th block is
\begin{align}
&\mathbf{y}_{j,\mathrm{b}}^{{\rm F}}[t]=\sqrt{\rho\alpha_{\mathrm{b}}}\mathbf{H}_{s}^{\nabla}\mathbf{x}_{i}[t]+\sqrt{\rho\alpha_{\mathrm{b}}}\mathbf{H}^{\triangle}_{s}\mathbf{x}_{i}[t-1]+\sqrt{\rho\alpha_{\mathrm{c}}}\mathbf{x}_{j}[t]\notag
\\&+\sqrt{\rho\alpha_{\mathrm{m}}}\mathbf{H}_{j}^{m\nabla}\mathbf{x}_{j}[t]+\sqrt{\rho\alpha_{\mathrm{m}}}\mathbf{H}^{m\triangle}_{j}\mathbf{x}_{j}[t-1]+\mathbf{w}_{j,\mathrm{b}}[t],
\label{Eq:Down_Backhaul_Received_Signal}
\end{align}
where $\sqrt{\alpha_{\mathrm{c}}}\mathbf{x}_{j}$ represents the signal leakage from the circulator of the FD radio, $\alpha_{\mathrm{m}}$ is the path loss factor modeling the attenuation affecting the component of the SI that experiences a multi-path propagation, and $\alpha_{\mathrm{b}}$ is the path loss factor characterizing the link between the two SNs. Given the nature of the communication, we can safely assume a block-wise decoding of the signaling at each SN. In this context, a successive interference cancellation approach could be adopted to correctly subtract the signal decoded in one block from the signal present in the subsequent block. Thus, if a perfect cancellation can be performed, the obtained signal after the SIC can be written as
\begin{align}
&\mathbf{y}_{j,\mathrm{b}}^{{\rm SIC}}[t]=\sqrt{\rho\alpha_{\mathrm{b}}}\mathbf{H}_{s}^{\nabla}\mathbf{\Gamma}_{i}^{\rm F}\mathbf{C}_{i}^{\rm F}\mathbf{u}_{i,\mathrm{b}}[t]+\mathbf{w}_{j,\mathrm{b}}[t]\notag
\\&\quad+\sqrt{\rho\alpha_{\mathrm{b}}}\left(\mathbf{H}_{s}^{\nabla}\mathbf{A}\mathbf{F}^{-1}\mathbf{u}_{i,\mathrm{o}}[t]+\mathbf{H}^{\triangle}_{s}\mathbf{A}\mathbf{F}^{-1}\mathbf{u}_{i,\mathrm{o}}[t-1]\right)\notag
\\ &\qquad +\sqrt{\rho\alpha_{\mathrm{eq}}}\left(\mathbf{A}\mathbf{F}^{-1}\mathbf{u}_{j,\mathrm{o}}[t]+\mathbf{\Gamma}_{j}^{\rm F}\mathbf{C}_{j}^{\rm F}\mathbf{u}_{j,\mathrm{b}}[t]\right),\label{Eq:Down_Backhaul_Processed_Signal_Decoding}
\end{align}
with $\mathbf{w}_{j,\mathrm{b}}\in\mathcal{CN}(\mathbf{0},N_{0}\mathbf{I}_{N+L})$ an AWGN and where
\begin{align}
\alpha_{\mathrm{eq}}&=\begin{cases}
0,&\mbox{~if~} \rho\leq\frac{P_{\mathrm{th}}}{P},\\
\frac{N_0}{P_{\mathrm{th}}},&\mbox{~if~} \frac{P_{\mathrm{th}}}{P}<\rho\leq\min\left\{1,\frac{P_{\mathrm{sat}}}{P}\right\}.
\end{cases}\label{Eq:Equiv_Leak_Interference_Path_Loss} 
\end{align}
The power of the multi-path SI signal in practical implementations is typically at least $25$~dB lower than the power of the SI signal leaking from the circulator\cite{conf:knox12}. In practice, it can be safely assumed that the contribution of the former to the overall SI signal in \eqref{Eq:Down_Backhaul_Processed_Signal_Decoding} is quantitatively negligible. Then, only the SI signal leaking from the circulator will be considered in the following, for simplicity. At this stage, it is convenient to divide the analysis into two parts, for the sake of completeness. First, we will focus on the case $\rho\leq\frac{P_{\mathrm{th}}}{P}$, in which no residual SI affects the decoding of the signaling. Then, we will study the case $\frac{P_{\mathrm{th}}}{P}<\rho\leq\min\left\{1,\frac{P_{\mathrm{sat}}}{P}\right\}$, in which residual SI remains after the SIC.

\subsubsection{Absence of Residual Self-interference}\label{Subsubsec: Down_Without_Residual_Self-interference}

From \eqref{Eq:Equiv_Leak_Interference_Path_Loss} we have that $\alpha_{\mathrm{eq}}=0$ in this case, then \eqref{Eq:Down_Backhaul_Processed_Signal_Decoding} can be rewritten as
\begin{align}
&\mathbf{y}_{j,\mathrm{b}}^{{\rm SIC}}[t]=\sqrt{\rho\alpha_{\mathrm{b}}}\mathbf{H}_{s}^{\nabla}\mathbf{\Gamma}_{i}^{{\rm F}}\mathbf{C}_{i}^{{\rm F}}\mathbf{u}_{i,\mathrm{b}}[t]
+\mathbf{z}_{j,\mathrm{b}}[t],
\label{Eq:Down_Without_Backhaul_Processed_Signal_Decoding}
\end{align}
where $\mathbf{z}_{j,\mathrm{b}}[t]=\mathbf{w}_{j,\mathrm{b}}[t]+\sqrt{\rho\alpha_{\mathrm{b}}}\big(\mathbf{H}_{s}^{\nabla}\mathbf{A}\mathbf{F}^{-1}\mathbf{u}_{i,\mathrm{o}}[t]
+\mathbf{H}_{s}^{\triangle}\mathbf{A}\mathbf{F}^{-1}\mathbf{u}_{i,\mathrm{o}}[t-1]\big)$ is the equivalent noise. In this regard, we note that it is reasonable to assume that each SN may have an estimation of the covariance matrix of the OFDM signal transmitted by the other SN in this phase~\cite{art:speth99, art:tang07}. Accordingly, let 
\begin{align}
\mbox{cov}\big(\mathbf{z}_{j,\mathrm{b}}\big)&=N_{0}\mathbf{I}_{N+L}+\rho\alpha_{\mathrm{b}}\mathbf{H}_{s}^{\nabla}\mathbf{A}\mathbf{F}^{-1}\mathbf{P}^{{\rm F}}_{i,\mathrm{o}}\mathbf{F}\mathbf{A}^{\mathrm{H}}\big(\mathbf{H}_{s}^{\nabla}\big)^{\mathrm{H}}\notag
\\&\quad+\rho\alpha_{\mathrm{b}}\mathbf{H}_{s}^{\triangle}\mathbf{A}\mathbf{F}^{-1}\mathbf{P}^{{\rm F}}_{i,\mathrm{o}}\mathbf{F}\mathbf{A}^{\mathrm{H}}\big(\mathbf{H}_{s}^{\triangle}\big)^{\mathrm{H}}\label{Eq:Down_Without_Backhaul_EqivNoise_Cov}
\end{align} 
%
%
%
be the corresponding noise-whitening matrix. The whitened version of \eqref{Eq:Down_Without_Backhaul_Processed_Signal_Decoding} can be written as
%
%
\begin{align}
\mathbf{y}_{j,\mathrm{b}}^{\mathrm{SIC}}[t]&=\sqrt{\rho\alpha_{\mathrm{b}}}\mbox{cov}^{-\frac{1}{2}}\big(\mathbf{z}_{j,\mathrm{b}}\big)\mathbf{H}_{s}^{\nabla}\mathbf{\Gamma}_{i}^{{\rm F}}\mathbf{C}_{i}^{{\rm F}}\mathbf{u}_{i,\mathrm{b}}[t]\notag
\\&\quad+\mbox{cov}^{-\frac{1}{2}}\big(\mathbf{z}_{j,\mathrm{b}}\big)\mathbf{z}_{j,\mathrm{b}}[t].
\label{Eq:Down_Without_Backhaul_Whitened_Signal}
\end{align}
Now, we focus on the first right-hand side element of \eqref{Eq:Down_Without_Backhaul_Whitened_Signal} and let $\mbox{cov}^{-\frac{1}{2}}\big(\mathbf{z}_{j,\mathrm{b}}\big)\mathbf{H}_{s}^{\nabla}\mathbf{\Gamma}_{i}^{{\rm F}}
=\mathbf{U}_{i}^{{\rm F}}\mathbf{\Sigma}_{i}^{{\rm F}}\big(\mathbf{Q}_{i}^{{\rm F}}\big)^{\mathrm{H}}$ be a singular value decomposition (SVD), with $\mathbf{U}_{i}^{{\rm F}}\in\mathbb{C}^{(N+L)\times(N+L)}$ and $\mathbf{Q}_{i}^{{\rm F}}\in\mathbb{C}^{L \times L}$ unitary matrices, and $\mathbf{\Sigma}_{i}^{{\rm F}}=[\mbox{diag}(\sigma_{1,\mathrm{z}}^{{\rm F}}, \dots, \sigma_{L,\mathrm{z}}^{{\rm F}}) \hspace{2mm} \mathbf{0}_{L \times N}]^\mathrm{T}$ where $\sigma_{i,\mathrm{z}}^{{\rm F}}$ is the $i$th singular value of $\mbox{cov}^{-\frac{1}{2}}\big(\mathbf{z}_{j,\mathrm{b}}\big)\mathbf{H}_{s}^{\nabla}\mathbf{\Gamma}_{i}^{{\rm F}}$. The link between the two SNs can then be decomposed into $L$ parallel independent Gaussian channels by letting $\mathbf{C}_{i}^{{\rm F}}=\mathbf{Q}_{i}^{{\rm F}}$ and defining $\big(\mathbf{U}_{i}^{{\rm F}}\big)^{\mathrm{H}}\mbox{cov}^{-\frac{1}{2}}\big(\mathbf{z}_{j,\mathrm{b}}\big)$ as the decoding matrix for the $i$th signaling. The achievable rate over the link between the two SNs in the forward phase, from the point of view of the $i$th SN, is then given by
\begin{align}
R^{{{\rm F}}}_{i,\mathrm{b}}\left(\rho\right)=\frac{1}{N+L}\sum_{n=1}^{L}\log_2\left(1+\rho\alpha_{\mathrm{b}} \big[\mathbf{P}^{{\rm F}}_{i,\mathrm{b}}\big]_{nn}\big[\mathbf{\Sigma}_{i}^{{\rm F}}\big]^2_{nn}\right),\label{Eq:Down_Without_Backhaul_Rate}
\end{align}
with
%
%
$\big[\mathbf{P}_{i,\mathrm{b}}^{{\rm F}}\big]_{nn}=\big(\kappa_{2,i}-\big(\rho\alpha_{\mathrm{b}}\big[\mathbf{\Sigma}_{i}^{{\rm F}}\big]^2_{nn}\big)^{-1}\big)^{+},\,
\forall\,n\in\{1,\cdots,L\}$
%
%
being the WF solution, where $\kappa_{2,i}$ is chosen such that $\mbox{Tr}\big(\mathbf{P}_{i,\mathrm{b}}^{{\rm F}}\big)=\left(N+L\right)P_{\mathrm{b}}$. In practice, the information required at both SNs to perform these computations could be exchanged in a preliminary phase of the communication, at the beginning of each new coherence time of the channel.

\subsubsection{Presence of Residual Self-interference}\label{Subsubsec: Down_With_Residual_Self-interference}

From \eqref{Eq:Equiv_Leak_Interference_Path_Loss} we have that $\alpha_{\mathrm{eq}}=\frac{N_0}{P_{\mathrm{th}}}$ in this case, then \eqref{Eq:Down_Backhaul_Processed_Signal_Decoding} can be rewritten as
\begin{align}
&\mathbf{y}_{j,\mathrm{b}}^{{\rm SIC}}[t]=\sqrt{\rho\alpha_{\mathrm{b}}}\mathbf{H}_{s}^{\nabla}\mathbf{\Gamma}_{i}^{{\rm F}}\mathbf{C}_{i}^{{\rm F}}\mathbf{u}_{i,\mathrm{b}}[t]
+\mathbf{z}_{j,\mathrm{b}}[t],
\label{Eq:Down_With_Backhaul_Processed_Signal_Decoding}
\end{align}
where
\begin{align}
&\mathbf{z}_{j,\mathrm{b}}[t]=\mathbf{w}_{j,\mathrm{b}}[t]+\sqrt{\frac{\rho N_0}{P_{\mathrm{th}}}}\left(\mathbf{A}\mathbf{F}^{-1}\mathbf{u}_{j,\mathrm{o}}[t]+\mathbf{\Gamma}_{j}^{{\rm F}}\mathbf{C}_{j}^{{\rm F}}\mathbf{u}_{j,\mathrm{b}}[t]\right)\notag
\\&+\sqrt{\rho\alpha_{\mathrm{b}}}\left(\mathbf{H}_{s}^{\nabla}\mathbf{A}\mathbf{F}^{-1}\mathbf{u}_{i,\mathrm{o}}[t]+\mathbf{H}^{\triangle}_{s}\mathbf{A}\mathbf{F}^{-1}\mathbf{u}_{i,\mathrm{o}}[t-1]\right)
\label{Eq:Down_With_Backhaul_Eqiv_Noise}
\end{align}
is the equivalent noise. Following the same procedures as for the previous case we write $\mbox{cov}^{-\frac{1}{2}}\big(\mathbf{z}_{j,\mathrm{b}}\big)\mathbf{H}_{s}^{\nabla}\mathbf{\Gamma}_{i}^{{\rm F}}
=\mathbf{U}_{i}^{{\rm F}}\mathbf{\Sigma}_{i}^{{\rm F}}\big(\mathbf{Q}_{i}^{{\rm F}}\big)^{\mathrm{H}}$, where $\mathbf{U}_{i}^{{\rm F}}$,  $\mathbf{\Sigma}_{i}^{{\rm F}}$, and $\mathbf{Q}_{i}^{{\rm F}}$ are defined analogously. Computing $\mbox{cov}\big(\mathbf{z}_{j,\mathrm{b}}\big)$ is rather complex in this case. Differently from before, herein the adoption of an adequate information exchange protocol between the SNs would not be sufficient for the latter to compute this quantity. In fact, no prior information about $\mathbf{C}_{j}^{{\rm F}}$ is available at the $i$th SN. Furthermore, the power allocation for $\mathbf{u}_{j,\mathrm{b}}[t]$ at the $j$th SN impacts both the power profile of the SI affecting the $j$th SN itself and the spectral efficiency of the link towards the $i$th SN. In practice, these two phenomena are strongly coupled. 

To cope with this issue, iterative WF algorithms should be performed at both SNs to maximize the spectral efficiency of the link between them. However, the coherence time of the channels towards the ANs (upon which the CIA precoder is based) imposes strict time requirements to the SNs. Thus, iterative approaches are not feasible, especially in the absence of a wired connection between the SNs, and alternative strategies need to be devised. Interestingly, these two issues could be addressed by resorting to two simplifications. First, we conjecture that $\mathbf{C}_{j}^{{\rm F}}$ is a unitary matrix, as per the derivation performed for the previous case. Additionally, we assume a uniform power allocation for the $L$ symbols of $\mathbf{u}_{j,\mathrm{b}}[t]$, such that $\mathbf{P}_{j,\mathrm{b}}^{{\rm F}}=\frac{(N+L)P_{\mathrm{b}}}{L}\mathbf{I}_{L}$. An approximation of the noise-whitening matrix in this case can then be obtained as
\begin{align}
&\mbox{cov}\big(\mathbf{z}_{j,\mathrm{b}}\big)=N_{0}\mathbf{I}_{N+L}+\rho\alpha_{\mathrm{b}}\mathbf{H}_{s}^{\nabla}\mathbf{A}\mathbf{F}^{-1}\mathbf{P}^{{\rm F}}_{i,\mathrm{o}}\mathbf{F}\mathbf{A}^{\mathrm{H}}\big(\mathbf{H}_{s}^{\nabla}\big)^{\mathrm{H}}\notag
\\&+\rho\alpha_{\mathrm{b}}\mathbf{H}_{s}^{\triangle}\mathbf{A}\mathbf{F}^{-1}\mathbf{P}^{{\rm F}}_{i,\mathrm{o}}\mathbf{F}\mathbf{A}^{\mathrm{H}}\big(\mathbf{H}_{s}^{\triangle}\big)^{\mathrm{H}}+\frac{\rho N_0}{P_{\mathrm{th}}}\mathbf{A}\mathbf{F}^{-1}\mathbf{P}^{{\rm F}}_{j,\mathrm{o}}\mathbf{F}\mathbf{A}^{\mathrm{H}}\notag
\\&\qquad+\frac{\rho N_0\left(N+L\right)P_{\mathrm{b}}}{L P_{\mathrm{th}}}\mathbf{\Gamma}_{j}^{{\rm F}}\big(\mathbf{\Gamma}_{j}^{{\rm F}}\big)^{\mathrm{H}}.
\label{Eq:Down_With_Backhaul_EqivNoise_Cov}
\end{align}
Now, we let $\mathbf{C}_{i}^{{\rm F}}=\mathbf{Q}_{i}^{{\rm F}}$ and $\big(\mathbf{U}_{i}^{{\rm F}}\big)^{\mathrm{H}}\mbox{cov}^{-\frac{1}{2}}\big(\mathbf{z}_{j,\mathrm{b}}\big)$ be the decoding matrix for the signaling and, similarly to the previous case, we obtain  the achievable rate over the link between the two SNs in the forward phase as
\begin{align}
R^{{\rm F}}_{i,\mathrm{b}}\left(\rho\right)&=\frac{1}{N+L}\sum_{n=1}^{L}\log_2\left(1+\frac{\rho\alpha_{\mathrm{b}} \left(N+L\right)P_{\mathrm{b}}\big[\mathbf{\Sigma}_{i}^{{\rm F}}\big]^2_{nn}}{L}\right).\label{Eq:Down_With_Backhaul_Rate}
\end{align}
%
%
%
\subsection{The Impact of the Splitting Ratio}

By looking at \eqref{Eq:Down_Without_Backhaul_Rate} and \eqref{Eq:Down_With_Backhaul_Rate} we see that $\rho$ is present in the in-log term both explicitly and implicitly via $\big[\mathbf{P}_{i,\mathrm{b}}^{{\rm F}}\big]_{nn}$ and $\big[\mathbf{\Sigma}_{i}^{{\rm F}}\big]^2_{nn}$. A closed-form analysis of its impact on \eqref{Eq:Down_Without_Backhaul_Rate} and \eqref{Eq:Down_With_Backhaul_Rate} is not straightforward. However, we can address this problem as follows. We first define $\rho^*$ as the value at which the rate of the incoming signaling is maximum, i.e., the optimal splitting ratio. Subsequently, we note that the power of both the received signaling and interference signals increases with $\rho$ at the same pace, by construction, whereas $N_{0}$ is independent of the latter parameter. Thus, we can intuitively infer that both the SINR and achievable rate over the link between the two SNs are monotonically increasing functions of $\rho$ (the latter being a logarithmic function of the splitting ratio), for $0\leq\rho\leq\rho^*$. Consequently, the optimal splitting ratio is deterministically given by $\rho^*=\frac{P_{\mathrm{th}}}{P}$ or $\rho^*=\min\left\{1,\frac{P_{\mathrm{sat}}}{P}\right\}$ if residual SI is absent or present, respectively. We note that these deductions will be verified in Sec~\ref{Sec: Numerical results}. At this stage, we can safely assume a priori that the curve representing the achievable rate as a function of the SINR may be approximated by $g(\rho,k)=\log_2\left(1+c_k\rho^{\phi_k}\right)$, where $c_k$ and $\phi_k$ depend on both the concavity of the curve and on the absence (e.g., $k=1$) or presence (i.e., $k=2$) of residual SI. In practice, $c_k$ and $\phi_k$ can be identified by the SNs following a simple one-shot procedure. After the first CSI has been acquired by the SNs, an estimation of the instantaneous achievable rate is computed using \eqref{Eq:Down_Without_Backhaul_Rate} or \eqref{Eq:Down_With_Backhaul_Rate}, for two values of $\rho$ arbitrarily chosen. For instance, reasonable choices for the interval $0\leq\rho\leq\frac{P_{\mathrm{th}}}{P}$, i.e., the absence of residual SI, are $\rho=\frac{P_{\mathrm{th}}}{P}$ and $\rho=\frac{P_{\mathrm{th}}}{2P}$. Similarly,  reasonable choices within the interval $\frac{P_{\mathrm{th}}}{P}<\rho\leq\min\left\{1,\frac{P_{\mathrm{sat}}}{P}\right\}$ could be $\rho=\frac{P_{\mathrm{th}}+\epsilon}{P}$, with $\epsilon\to 0$, and $\rho=\min\left\{1,\frac{P_{\mathrm{sat}}}{P}\right\}$. The following proposition holds.

%
%
\begin{proposition}\label{Lem:Down_Lemma_Backhaul}
	The achievable rate over the link between the $i$th and the $j$th SN in the forward phase can be approximated as 
	\begin{align}
	R_{i,\mathrm{b}}^{{\rm F}}\left(\rho\right)&\approx\begin{cases}
	\log_2\left(1+c_1\rho^{\phi_1}\right),\mbox{~when~}\rho\leq \frac{P_{\mathrm{th}}}{P},\\
	\log_2\left(1+c_2\rho^{\phi_2}\right),\\
	\hspace{2em}\mbox{when~}\frac{P_{\mathrm{th}}}{P}<\rho\leq\min\left\{1,\frac{P_{\mathrm{sat}}}{P}\right\},
	\end{cases}\label{Eq:Lem_1_first_case}
	\end{align}
	where
	\begin{align*}
	c_1&=\left(2^{ R_{i,\mathrm{b}}^{{\rm F}}(\frac{P_{\mathrm{th}}}{P})}-1\right)\left(\frac{P}{P_{\mathrm{th}}}\right)^{\phi_1}, 
\\	\phi_1&=\log_2\left(\frac{2^{ R_{i,\mathrm{b}}^{{\rm F}}(\frac{P_{\mathrm{th}}}{P})}-1}{2^{ R_{i,\mathrm{b}}^{{\rm F}}(\frac{P_{\mathrm{th}}}{2P})}-1}\right),
\\c_2&=\left(2^{ R_{i,\mathrm{b}}^{{\rm F}}(\frac{P_{\mathrm{th}}+\epsilon}{P})}-1\right)\left(\frac{P}{P_{\mathrm{th}}+\epsilon}\right)^{\phi_2}, 
\end{align*}
\begin{align*}
	\phi_2&=\log_{\frac{P_{\mathrm{th}}+\epsilon}{\min\{P,P_{\mathrm{sat}}\}}}\left(\frac{2^{ R_{i,\mathrm{b}}^{{\rm F}}(\frac{P_{\mathrm{th}}+\epsilon}{P})}-1}{2^{ R_{i,\mathrm{b}}^{{\rm F}}(\min\{1,\frac{P_{\mathrm{sat}}}{P}\})}-1}\right).
	\end{align*}
\end{proposition}
\begin{IEEEproof}
The result in \eqref{Eq:Lem_1_first_case}, can be obtained by enforcing a perfect match between the aforementioned rate estimation and $g(\rho, k)$. Thus, in the case of the absence of residual SI in the RX chain, i.e., $\rho\leq \frac{P_{\mathrm{th}}}{P}$, we have
\begin{align*}
\begin{cases}R_{i,\mathrm{b}}^{{\rm F}}\left(\frac{P_{\mathrm{th}}}{P}\right)=\log_2\left(1+c_1\left(\frac{P_{\mathrm{th}}}{P}\right)^{\phi_1}\right),
\\R_{i,\mathrm{b}}^{{\rm F}}\left(\frac{P_{\mathrm{th}}}{2P}\right)=\log_2\left(1+c_1\left(\frac{P_{\mathrm{th}}}{2P}\right)^{\phi_1}\right).
\end{cases}
\end{align*}
Then, the two parameters $c_1$ and $\phi_1$ can be straightforwardly obtained as
\begin{align*}
	c_1&=\left(2^{ R_{i,\mathrm{b}}^{{\rm F}}(\frac{P_{\mathrm{th}}}{P})}-1\right)\left(\frac{P}{P_{\mathrm{th}}}\right)^{\phi_1},
	\\\phi_1&=\log_2\left(\frac{2^{ R_{i,\mathrm{b}}^{{\rm F}}(\frac{P_{\mathrm{th}}}{P})}-1}{2^{ R_{i,\mathrm{b}}^{{\rm F}}(\frac{P_{\mathrm{th}}}{2P})}-1}\right).
\end{align*}
The two parameters $c_2$ and $\phi_2$ can be computed analogously, in the case of the presence of residual SI in the RX chain, i.e., when $\frac{P_{\mathrm{th}}}{P}<\rho\leq\min\left\{1,\frac{P_{\mathrm{sat}}}{P}\right\}$.
\end{IEEEproof}
Proposition~\ref{Lem:Down_Lemma_Backhaul} serves a two-fold purpose. On one hand, it provides an approximation of the achievable rate over the link between the two SNs in the forward phase. On the other hand, it allows to identify the values of $\rho$ for which the proposed FD architecture can outperform the state-of-the-art solutions in terms of throughput. In this sense, we recall that the performance of the state of the art can be easily obtained by setting $\rho=1$, i.e., virtually deactivating the 3-port composite element described in Sec.~\ref{Sec: FD architecture}. Now, consider the relevant case $P_{\mathrm{th}}<P\leq P_{\mathrm{sat}}$, i.e., the FD device implemented according to state-of-the-art solutions suffers from residual SI but no saturation occurs in its RX chain. The following corollary directly descends from Proposition~\ref{Lem:Down_Lemma_Backhaul}. 
%
%
\begin{corollary}\label{Lem:Down_Remark_Backhaul}
If $P_{\mathrm{th}}<P\leq P_{\mathrm{sat}}$ in the forward phase, then the proposed FD architecture delivers a higher throughput than the state-of-the-art FD solutions, for $\rho\in\bigg[\sqrt[\phi_1]{\frac{2^{R^{{\rm F}}_{i,\mathrm{b}}\left(1\right)}-1}{c_1}},\frac{P_{\mathrm{th}}}{P}\bigg]$.
\end{corollary}
%
%
\begin{IEEEproof}
When $\frac{P_{\mathrm{th}}}{P}<\rho\leq 1$, the approximated achievable rate is a monotonically increasing function. The maximum rate is then achieved for $\rho=1$, i.e., the state-of-the-art solution is always better in this region. Alternatively, when $\rho\leq \frac{P_{\mathrm{th}}}{P}$, we let $\log_2\left(1+c_1\rho^{\phi_1}\right)\geq R^{{\rm F}}_{i,\mathrm{b}}(1)$, as per Proposition~\ref{Lem:Down_Lemma_Backhaul}, and the result of Corollary~\ref{Lem:Down_Remark_Backhaul} can be directly obtained.
\end{IEEEproof}
%
%
\subsection{Recycled Energy} \label{sec:energy_forward}

A seen in Sec.~\ref{Sec: FD architecture}, the 3-port composite element present in the proposed FD architecture includes an EH that can recycle a portion of the leaked RF energy in the form of DC energy. Using the digital representation of the signals for the sake of compactness of the notation, let $E_{i}^{{\rm F}}\left(\rho\right)$ be the recycled energy per symbol by the 3-port element at the $i$th SN, in the forward phase. By some straightforward derivations, this quantity can be computed as
\begin{align}
&E_{i}^{{\rm F}}\left(\rho\right)=\frac{\beta\left(1-\rho\right)}{N+L}\times\mbox{Tr}\Big(\alpha_{\mathrm{b}}\left(\big(\mathbf{H}_{s}^{\nabla}\big)^{\mathrm{H}}\mathbf{H}_{s}^{\nabla}+\big(\mathbf{H}_{s}^{\triangle}\big)^{\mathrm{H}}\mathbf{H}_{s}^{\triangle}\right)\notag
\\&\times\left(\mathbf{A}\mathbf{F}^{-1}\mathbf{P}^{{\rm F}}_{j,\mathrm{o}}\mathbf{F}\mathbf{A}^{\mathrm{H}}+\mathbf{\Gamma}_{j}^{{\rm F}}\mathbf{C}_{j}^{{\rm F}}\mathbf{P}^{{\rm F}}_{j,\mathrm{b}}\big(\mathbf{C}_{j}^{{\rm F}}\big)^{\mathrm{H}}\big(\mathbf{\Gamma}_{j}^{{\rm F}}\big)^{\mathrm{H}}\right)\notag
\\&+\Big(\alpha_{\mathrm{m}}\big(\mathbf{H}_{i}^{m\triangle}\big)^{\mathrm{H}}\mathbf{H}_{i}^{m\triangle}+\left(\sqrt{\alpha_{\mathrm{c}}}\mathbf{I}_{N+L}+\sqrt{\alpha_{\mathrm{m}}}\mathbf{H}_{i}^{m\nabla}\right)^{\mathrm{H}}\notag
\\&\qquad\times\left(\sqrt{\alpha_{\mathrm{c}}}\mathbf{I}_{N+L}+\sqrt{\alpha_{\mathrm{m}}}\mathbf{H}_{i}^{m\nabla}\right)\Big)\notag
\\&\times\left(\mathbf{A}\mathbf{F}^{-1}\mathbf{P}^{{\rm F}}_{i,\mathrm{o}}\mathbf{F}\mathbf{A}^{\mathrm{H}}+\mathbf{\Gamma}_{i}^{{\rm F}}\mathbf{C}_{i}^{{\rm F}}\mathbf{P}^{{\rm F}}_{i,\mathrm{b}}\big(\mathbf{C}_{i}^{{\rm F}}\big)^{\mathrm{H}}\big(\mathbf{\Gamma}_{i}^{{\rm F}}\big)^{\mathrm{H}}\right)\Big)
,\label{Eq:Down_Harvested_Energy_at_SN}
\end{align}
with $\beta$ as the RF-to-DC energy conversion efficiency~\cite{art:zhang13}. At this stage, it is worth introducing a useful approximation of \eqref{Eq:Down_Harvested_Energy_at_SN}, and increasing its qualitative descriptiveness. Thus, we neglect the received signaling and multi-path SI signal terms from \eqref{Eq:Down_Harvested_Energy_at_SN}, given their different order of magnitude as compared of the intensity of the SI~\cite{conf:knox12}, and approximate the recycled energy per symbol at the $i$th SN as $E_{i}^{{\rm F}}\left(\rho\right)\approx \beta\alpha_{\mathrm{c}} \left(1-\rho\right)P$.

\section{The Backward Phase}\label{Sec: Uplink}

We switch now our focus to the backward phase.
\subsection{Signal Model}

Similar to the forward phase, the ANs communicate with the SNs by means of an OFDMA transmission. The information transmitted by the $i$th AN in the $t$th OFDM block is given by 
\begin{equation}
\big[\mathbf{v}_{i,\mathrm{o}}[t]\big]_n=\begin{cases}
\big[\mathbf{v}_{i,\mathrm{o}}[t]\big]_n\in\mathbb{C},&\hspace{.5mm}n\in \mathscr{N}_i,\\
0, &\mathrm{otherwise},
\end{cases}
\label{Eq:Up_OFDM_Information}
\end{equation}
with covariance matrix $\mathbf{P}^{{\rm B}}_{i,\mathrm{o}}\in\mathbb{C}^{N\times N}$. The $i$th transmitted signal in the backward phase reads $\mathbf{s}_{i,\mathrm{o}} [t]=\mathbf{A}\mathbf{F}^{-1}\mathbf{v}_{i,\mathrm{o}}[t]$, with $\mbox{Tr}\big(\mathbb{E}[\mathbf{s}_{i,\mathrm{o}}[t]\big(\mathbf{s}_{i,\mathrm{o}}[t]\big)^{\mathrm{H}}]\big)=(N+L)P_{A}$, where $P_{A}$ is the TX power budget at the AN. We recall that the ANs are not the only devices active in this phase of communication. In fact, the FD SNs are equally active and exchange signaling while receiving the backward signals from the ANs. We denote the signaling transmitted by the $i$th SN in the $t$th block as $\mathbf{s}_{i,\mathrm{b}}[t]\in\mathbb{C}^{N+L}$, with $\mbox{Tr}\big(\mathbb{E}[\mathbf{s}_{i,\mathrm{b}}[t]\big(\mathbf{s}_{i,\mathrm{b}}[t]\big)^{\mathrm{H}}]\big)=(N+L)P$.

At this stage, we recall that the SNs do not transmit information to the ANs during the backward phase. Thus after the operations performed by the 3-port composite element, the overall received signal at the $i$th SN, including both the SI and the signals from the other devices, is
\begin{align}
&\mathbf{y}_{i}[t]=\sqrt{\rho}\Big(\sqrt{\alpha_{ii}}\mathbf{H}_{ii}^{\nabla}\mathbf{s}_{i,\mathrm{o}}[t]+\sqrt{\alpha_{ii}}\mathbf{H}^{\triangle}_{ii}\mathbf{s}_{i,\mathrm{o}}[t-1]\notag
\\&+\sqrt{\alpha_{ij}}\mathbf{H}_{ij}^{\nabla}\mathbf{s}_{j,\mathrm{o}}[t]+\sqrt{\alpha_{ij}}\mathbf{H}^{\triangle}_{ij}\mathbf{s}_{j,\mathrm{o}}[t-1]+\sqrt{\alpha_{\mathrm{c}}}\mathbf{s}_{i,\mathrm{b}}[t]\notag 
\\&+\sqrt{\alpha_{\mathrm{b}}}\mathbf{H}_{s}^{\nabla}\mathbf{s}_{j,\mathrm{b}}[t]+\sqrt{\alpha_{\mathrm{b}}}\mathbf{H}^{\triangle}_{s}\mathbf{s}_{j,\mathrm{b}}[t-1]\notag
\\&+\sqrt{\alpha_{\mathrm{m}}}\mathbf{H}_{i}^{m\nabla}\mathbf{s}_{i,\mathrm{b}}[t]+\sqrt{\alpha_{\mathrm{m}}}\mathbf{H}^{m\triangle}_{i}\mathbf{s}_{i,\mathrm{b}}[t-1]\Big)+\mathbf{w}_{i}[t].
\label{Eq:P2_Received_Signal}
\end{align}
Subsequently, after cancelling the SI, the information signal to be decoded by the $i$th SN reads
\begin{align}
&\mathbf{y}_{i}^{{\rm SIC}}[t]=\sqrt{\rho}\Big(\sqrt{\alpha_{ii}}\mathbf{H}_{ii}^{\nabla}\mathbf{s}_{i,\mathrm{o}}[t]+\sqrt{\alpha_{ii}}\mathbf{H}^{\triangle}_{ii}\mathbf{s}_{i,\mathrm{o}}[t-1]\notag
\\&+\sqrt{\alpha_{ij}}\mathbf{H}_{ij}^{\nabla}\mathbf{s}_{j,\mathrm{o}}[t]+\sqrt{\alpha_{ij}}\mathbf{H}^{\triangle}_{ij}\mathbf{s}_{j,\mathrm{o}}[t-1]+\sqrt{\alpha_{\mathrm{eq}}}\mathbf{s}_{i,\mathrm{b}}[t]\notag
\\&\quad+\sqrt{\alpha_{\mathrm{b}}}\mathbf{H}_{s}^{\nabla}\mathbf{s}_{j,\mathrm{b}}[t]+\sqrt{\alpha_{\mathrm{b}}}\mathbf{H}^{\triangle}_{s}\mathbf{s}_{j,\mathrm{b}}[t-1]\Big)+\mathbf{w}_{i}[t],\label{Eq:P2_Split_Signal_Decoding}
\end{align}
where $\mathbf{w}_{i}\in\mathcal{CN}\left(\mathbf{0},N_{0}\mathbf{I}_{N+L}\right)$ is an AWGN. Now, first the backward signal is decoded by the $i$th SN by means of legacy OFDM demodulation. Similar to the previous case, a block-wise decoding of the backward signal is performed at each SN. A successive interference cancellation is then performed by subtracting the decoded backward signal from \eqref{Eq:P2_Split_Signal_Decoding} and finally the decoding of the signaling takes place. In this regard, we note that a perfect cancellation is assumed.

\subsection{Performance of the Transmission in the Backward Phase}\label{Subsubsec: Without Uplink OFDM}

As for \eqref{Eq:Down_Backhaul_Processed_Signal_Decoding}, residual SI can be absent or present in the RX chain, depending on the value of $\rho$. Thus, as before, the following analysis encompasses the two cases, for the sake of completeness. 

\subsubsection{Absence of Residual Self-interference}\label{Subsec: Without_Self-interference}

The only potential source of interference during the OFDM demodulation is given by $\mathbf{s}_{j,\mathrm{b}}[t]$. In this context, the $i$th SN designs its signaling in order not to induce a rate loss on the link between the $i$th SN/AN pair in the backward phase. We know from the discussion in Sec.~\ref{Sec: Downlink}, that if $\mathbf{s}_{i,\mathrm{b}}[t]$ satisfies
\begin{align}\label{Eq:P2_Backhaul_Constraint_Dumb}
\mathbf{D}_{j}\mathbf{F}\mathbf{B}\mathbf{H}_{s}^{\nabla}\mathbf{s}_{i,\mathrm{b}}[t]=\mathbf{0},
~i \neq j,
\end{align}
then its impact on the SINR of the backward signal at the $j$th SN is zeroed, and the decoding of the latter can proceed according to the conventional OFDMA scheme. Now, by plugging $\alpha_{\mathrm{eq}}=0$ into \eqref{Eq:P2_Split_Signal_Decoding}, we can write the demodulated signal in this case as
\begin{align}
\mathbf{y}_{i,\mathrm{o}}^{{\rm B}}[t]&=\sqrt{\rho\alpha_{ii}}\mathbf{D}_{i}\mbox{diag}\big(\mathbf{\tilde{h}}_{ii}\big)\mathbf{v}_{i,\mathrm{o}}[t]+\mathbf{D}_{i}\mathbf{F}\mathbf{B}\mathbf{w}_{i}[t].
\label{Eq:P2_Without_OFDM_WithoutIBI}
\end{align}
The achievable rate over the link between the $i$th SN/AN pair in this case is then 
\begin{align}\label{Eq:P2_Without_OFDM_Rate}
&R_{i,\mathrm{o}}^{{\rm B}}\left(\rho\right)=\frac{1}{N+L}\sum_{n\in\mathscr{N}_{i}}\log_2\left(1+\frac{\rho\alpha_{ii} [\mathbf{P}^{{\rm B}}_{i,\mathrm{o}}]_{nn}\lvert[\mathbf{\tilde{h}}_{ii}]_{n}\rvert^2}{N_{0}}\right),
\end{align}
where $\big[\mathbf{P}^{{\rm B}}_{i,\mathrm{o}}\big]_{nn}=\left(\kappa_{3,i}-\frac{N_{0}}{\rho\alpha_{ii}\lvert[\mathbf{\tilde{h}}_{ii}]_{n}\rvert^2}\right)^{+},~\forall\,n\in\mathscr{N}_{i}$ is the WF solution, with $\kappa_{3,i}$ chosen such that  $\mbox{Tr}\big(\mathbf{P}^{{\rm B}}_{i,\mathrm{o}}\big)=NP_{A}$. Furthermore, it is straightforward to see that $\rho^*=\frac{P_{\mathrm{th}}}{P}$ in this case.

\subsubsection{Presence of Residual Self-interference}\label{Subsecsec:Up_With_SI}

Provided that \eqref{Eq:P2_Backhaul_Constraint_Dumb} is satisfied, and plugging $\alpha_{\mathrm{eq}}=\frac{N_0}{P_{\mathrm{th}}}$ into \eqref{Eq:P2_Split_Signal_Decoding}, we can write
\begin{align}
\mathbf{y}_{i,\mathrm{o}}^{{\rm B}}[t]&=\sqrt{\rho\alpha_{ii}}\mathbf{D}_{i}\mbox{diag}\big(\mathbf{\tilde{h}}_{ii}\big)\mathbf{v}_{i,\mathrm{o}}[t]\notag
\\&\quad+\mathbf{D}_{i}\mathbf{F}\mathbf{B}\left(\mathbf{w}_{i}[t]+\sqrt{\frac{\rho N_0}{P_{\mathrm{th}}}}\mathbf{s}_{i,\mathrm{b}}[t]\right) .\label{Eq:P2_With_OFDM_Decoded}
\end{align}
The equivalent noise in \eqref{Eq:P2_With_OFDM_Decoded} is colored due to the presence of residual SI. However, a noise-whitening cannot be performed straightforwardly in this case. In fact, each $N$-sized vector $\mathbf{y}_{i,\mathrm{o}}^{{\rm B}}[t]$ includes $\left|\mathscr{N}_{j}\right|$ zero elements by construction. This issue can be addressed without loss of generality and correctness by removing the zero elements from $\mathbf{y}_{i,\mathrm{o}}^{{\rm B}}[t]$, to obtain a $\left|\mathscr{N}_{i}\right|$-sized vector offering higher analytical tractability. Thus, we first let $\mathbf{o}_{i}^{{\rm B}}=[o_{i}^{{\rm B}}(1),\dots,o_{i}^{{\rm B}}(|\mathscr{N}_i|)]$ be an ordered vector of indexes, such that $o_{i}^{{\rm B}}(k) \in \mathscr{N}_i$, $\forall$ natural $k \in [1,|\mathscr{N}_i|]$. Subsequently, we define a modified $(\left|\mathscr{N}_{i}\right|\times N)$-sized sub-carrier selection matrix $\mathbf{\tilde{D}}_i = [\mathbf{e}_{o_{i}^{{\rm B}}(1)} \cdots \mathbf{e}_{o_{i}^{{\rm B}}(|\mathscr{N}_i|)}]^{\text{H}}$, where $\mathbf{e}_{o_{i}^{{\rm B}}(j)}$ is the ${o_{i}^{{\rm B}}(j)}$th standard unit vector. Finally, we can rewrite \eqref{Eq:P2_With_OFDM_Decoded} as
\begin{align}\label{Eq:P2_With_OFDM_Decoded_shrunk}
\mathbf{\tilde{y}}_{i,\mathrm{o}}^{{\rm B}}[t]&=\sqrt{\rho\alpha_{ii}}\mathbf{\tilde{D}}_{i}\mbox{diag}\big(\mathbf{\tilde{h}}_{ii}\big)\big(\mathbf{\tilde{D}}_{i}\big)^{\mathrm{H}}\mathbf{\tilde{v}}_{i,\mathrm{o}}[t]+\mathbf{z}_{i}[t],
\end{align}
where we let $\mathbf{\tilde{v}}_{i,\mathrm{o}}[t]=\mathbf{\tilde{D}}_{i}\mathbf{v}_{i,\mathrm{o}}[t]$, for the sake of consistency of the notation, and define $\mathbf{z}_{i}[t]=\mathbf{\tilde{D}}_{i}\mathbf{F}\mathbf{B}\left(\mathbf{w}_{i}[t]+\sqrt{\frac{\rho N_0}{P_{\mathrm{th}}}}\mathbf{s}_{i,\mathrm{b}}[t]\right)$ as the equivalent noise. Now we can proceed as done for the forward phase and first write the noise-whitening matrix as
\begin{align}\label{Eq:P2_With_OFDM_NoiseCov}
\mbox{cov}\left(\mathbf{z}_{i}\right)&=N_0\mathbf{I}_{\left|\mathscr{N}_{i}\right|}+\frac{\rho N_0(N+L)P}{(N+L-\left|\mathscr{N}_{j}\right|)P_{\mathrm{th}}}\notag
\\&\quad\times\mathbf{\tilde{D}}_{i}\mathbf{F}\mathbf{B}\mathbf{\Gamma}_{i}^{{\rm B}}\big(\mathbf{\Gamma}_{i}^{{\rm B}}\big)^{\mathrm{H}}\mathbf{B}^{\mathrm{H}}\mathbf{F}^{-1}\big(\mathbf{\tilde{D}}_{i}\big)^{\mathrm{H}},
\end{align}
and then compute $\mbox{cov}^{-\frac{1}{2}}\left(\mathbf{z}_{i}\right)\mathbf{\tilde{D}}_{i}\mbox{diag}\big(\mathbf{\tilde{h}}_{ii}\big)\big(\mathbf{\tilde{D}}_{i}\big)^{\mathrm{H}}=\mathbf{\tilde{U}}_{i}\mathbf{\tilde{\Lambda}}_{i}\big(\mathbf{\tilde{U}}_{i}\big)^{\mathrm{H}}$, where $\mathbf{\tilde{U}}_{i}\in\mathbb{C}^{\left|\mathscr{N}_{i}\right|\times \left|\mathscr{N}_{i}\right|}$ is a unitary matrix and $\mathbf{\tilde{\Lambda}}_{i}=[\mbox{diag}(\tilde{\lambda}_{1,\mathrm{z}}, \dots, \tilde{\lambda}_{\left|\mathscr{N}_{i}\right|,\mathrm{z}})]$, where $\tilde{\lambda}_{i,\mathrm{z}}$ is the $i$th eigenvalue of $\mbox{cov}^{-\frac{1}{2}}\left(\mathbf{z}_{i}\right)\mathbf{\tilde{D}}_{i}\mbox{diag}\big(\mathbf{\tilde{h}}_{ii}\big)\big(\mathbf{\tilde{D}}_{i}\big)^{\mathrm{H}}$. Analogously, we can decompose the link between the $i$th SN/AN pair into $\lvert\mathscr{N}_{i}\rvert$ parallel Gaussian channels by  letting
$\mathbf{P}^{{\rm B}}_{i,\mathrm{o}}=\big(\mathbf{\tilde{D}}_{i}\big)^{\mathrm{H}}\mathbf{\tilde{U}}_{i}\mathbf{\tilde{P}}_{i,\mathrm{o}}^{{\rm B}}\big(\mathbf{\tilde{U}}_{i}\big)^{\mathrm{H}}\mathbf{\tilde{D}}_{i}$, and $\big[\mathbf{\tilde{P}}_{i,\mathrm{o}}^{{\rm B}}\big]_{nn}=\left(\kappa_{4,i}-\frac{1}{\rho\alpha_{ii}\lvert[\mathbf{\tilde{\Lambda}}_{i}]_{nn}\rvert^2}\right)^{+}$, $\forall n\in\left\{1,\cdots,\left|\mathscr{N}_{i}\right|\right\}$, with $\kappa_{4,i}$ chosen such that $\mbox{Tr}\big(\mathbf{\tilde{P}}_{i,\mathrm{o}}^{{\rm B}}\big)=NP_{A}$. Thus, the achievable rate over this link is
\begin{align}
&R_{i,\mathrm{o}}^{{\rm B}}\left(\rho\right)=\frac{1}{N+L}\sum_{n=1}^{\left|\mathscr{N}_{i}\right|}\log_2\left(1+\rho\alpha_{ii} \big[\mathbf{\tilde{P}}_{i,\mathrm{o}}^{{\rm B}}\big]_{nn}\lvert[\mathbf{\tilde{\Lambda}}_{i}]_{nn}\rvert^2\right).\label{Eq:P2_With_OFDM_Rate}
\end{align}
Intuitively, the values of $\rho^*$ for the backward phase are the same as for the forward phase. Moreover, as before, assessing the impact that $\rho$ may have on \eqref{Eq:P2_Without_OFDM_Rate} and \eqref{Eq:P2_With_OFDM_Rate}, when a WF strategy is adopted, is not straightforward. Thus, we can proceed as done for Proposition~\ref{Lem:Down_Lemma_Backhaul} and state the following. 
%
%
%
\begin{proposition}\label{Lem:Up_Lemma_OFDM}
	The achievable rate over the link between the $i$th SN/AN pair in the backward phase can be approximated as
	\begin{align}
	R_{i,\mathrm{o}}^{{\rm B}}\left(\rho\right)&\approx\begin{cases}
	\log_2\left(1+c_3\rho^{\phi_3}\right),\mbox{~when~} \rho\leq \frac{P_{\mathrm{th}}}{P},\\
	\log_2\left(1+c_4\rho^{\phi_4}\right),\\
	\hspace{2em}\mbox{when~} \frac{P_{\mathrm{th}}}{P}<\rho\leq\min\left\{1,\frac{P_{\mathrm{sat}}}{P}\right\},
	\end{cases}\label{Eq:P2_Without_OFDM_Approx_Rate}
	\end{align}
	where
	\begin{align*}
	c_3&=\left(2^{ R_{i,\mathrm{o}}^{{\rm B}}(\frac{P_{\mathrm{th}}}{P})}-1\right)\left(\frac{P}{P_{\mathrm{th}}}\right)^{\phi_3},
	\\\phi_3&=\log_2\left(\frac{2^{ R_{i,\mathrm{o}}^{{\rm B}}(\frac{P_{\mathrm{th}}}{P})}-1}{2^{ R_{i,\mathrm{o}}^{{\rm B}}(\frac{P_{\mathrm{th}}}{2P})}-1}\right),
	\\c_4&=\left(2^{ R_{i,\mathrm{o}}^{{\rm B}}(\frac{P_{\mathrm{th}}+\epsilon}{P})}-1\right)\left(\frac{P}{P_{\mathrm{th}}+\epsilon}\right)^{\phi_4}, 
	\\\phi_4&=\log_{\frac{P_{\mathrm{th}}+\epsilon}{\min\{P,P_{\mathrm{sat}}\}}}\left(\frac{2^{ R_{i,\mathrm{o}}^{{\rm B}}(\frac{P_{\mathrm{th}}+\epsilon}{P})}-1}{2^{ R_{i,\mathrm{o}}^{{\rm B}}(\min\{1,\frac{P_{\mathrm{sat}}}{P}\})}-1}\right).
	\end{align*}
\end{proposition}
\begin{IEEEproof}
The result can be obtained as done for Proposition~\ref{Lem:Down_Lemma_Backhaul}.
\end{IEEEproof}

%
%
We now focus as before on the relevant case $P_{\mathrm{th}}<P\leq P_{\mathrm{sat}}$, i.e., the FD device implemented according to state-of-the-art solutions suffers from residual SI but no saturation occurs in its RX chain. In this case, the following result directly descends from Proposition~\ref{Lem:Up_Lemma_OFDM}.
%
%
\begin{corollary}\label{Lem:Up_Remark_OFDM}
When $P_{\mathrm{th}}<P\leq P_{\mathrm{sat}}$, the interval of $\rho$ for which the achievable rate over the link between the $i$th SN/AN pair is higher if the proposed approach is adopted over the state-of-the-art solutions can be approximated as $\bigg[\sqrt[\phi_3]{\frac{2^{R^{{\rm B}}_{i,\mathrm{o}}\left(1\right)}-1}{c_3}},\frac{P_{\mathrm{th}}}{P}\bigg]$.
\end{corollary}
%
%
\begin{IEEEproof}
The result can be obtained as done for Corollary~\ref{Lem:Down_Remark_Backhaul}.
\end{IEEEproof}

\vspace{-.4cm}
\subsection{Performance of the Signaling Between SNs}\label{Subsubsec: Without Uplink Backhaul}

We start by noting that \eqref{Eq:P2_Backhaul_Constraint_Dumb} implies that
$\mbox{dim}\big(\mbox{null}\big(\mathbf{D}_{j}\mathbf{F}\mathbf{B}\mathbf{H}_{s}^{\nabla}\big)\big)=N+L-\lvert\mathscr{N}_j\rvert$. Hence, the $i$th SN can transmit up to $N+L-\lvert\mathscr{N}_j\rvert$ independent information streams for signaling in the backward phase. We let $\mathbf{v}_{i,\mathrm{b}}[t]\in\mathbb{C}^{N+L-\lvert\mathscr{N}_j\rvert}$, with covariance matrix $\mathbf{P}^{{\rm B}}_{i,\mathrm{b}}$, be the information symbol vector transmitted by the $i$th to the $j$th SN in this phase. Then, if we adopt the same representation as in \eqref{Eq:Down_Backhaul_Transmitted_Signal}, we can express the $i$th signaling as 
$\mathbf{s}_{i,\mathrm{b}}=\mathbf{\Gamma}_{i}^{{\rm B}}\mathbf{C}_{i}^{{\rm B}}\mathbf{v}_{i,\mathrm{b}}$, with $\mathbf{\Gamma}_{i}^{{\rm B}}\in\mathbb{C}^{(N+L)\times(N+L-\lvert\mathscr{N}_j\rvert)}$ an arbitrary semi-unitary matrix such that $\mathbf{D}_{j}\mathbf{F}\mathbf{B}\mathbf{H}_{s}^{\nabla}\mathbf{\Gamma}_{i}^{{\rm B}}=\mathbf{0}$. Now, assume a block-wise decoding of the signaling at the $j$th SN, as done as for the forward phase. Thus, the received signal at the $j$th SN after the SIC, i.e., $\mathbf{y}_{j}^{{\rm SIC}}[t]$, can be obtained from \eqref{Eq:P2_Split_Signal_Decoding} as
\begin{align}
&\mathbf{y}_{j}^{{\rm SIC}}[t]=\sqrt{\rho}\Big(\sqrt{\alpha_{\mathrm{b}}}\mathbf{H}_{s}^{\nabla}\mathbf{s}_{i,\mathrm{b}}[t]+\sqrt{\alpha_{\mathrm{eq}}}\mathbf{s}_{j,\mathrm{b}}[t]\notag
\\&\quad+\sqrt{\alpha_{ji}}\mathbf{H}_{ji}^{\nabla}\mathbf{s}_{i,\mathrm{o}}[t]+\sqrt{\alpha_{ji}}\mathbf{H}^{\triangle}_{ji}\mathbf{s}_{i,\mathrm{o}}[t-1]\Big)+\mathbf{w}_{j}[t].\label{Eq:P2_Split_Signal_Decoding_subtracting}
\end{align}
Following the same approach adopted so far, we divide the rest of the analysis into two parts: first, we will focus on the case $\rho\leq\frac{P_{\mathrm{th}}}{P}$, in which no residual SI affects the decoding of the signaling; then, we will study the case $\frac{P_{\mathrm{th}}}{P}<\rho\leq\min\left\{1,\frac{P_{\mathrm{sat}}}{P}\right\}$, in which residual SI remains after the SIC.

\subsubsection{Absence of Residual Self-interference}

From \eqref{Eq:Equiv_Leak_Interference_Path_Loss} we have that $\alpha_{\mathrm{eq}}=0$ in this case, then \eqref{Eq:P2_Split_Signal_Decoding_subtracting} can be rewritten as
\begin{align}\label{Eq:P2_Without_Backhaul_Decoded}
\mathbf{y}_{j}^{{\rm SIC}}[t]=\sqrt{\rho\alpha_{\mathrm{b}}}\mathbf{H}_{s}^{\nabla}\mathbf{\Gamma}_{i}^{{\rm B}}\mathbf{C}_{i}^{{\rm B}}\mathbf{v}_{i,\mathrm{b}}[t]+\mathbf{z}_{j}[t],
\end{align}
with $\mathbf{z}_{j}[t]=\mathbf{w}_{j}[t]+\sqrt{\rho\alpha_{ji}}\big(\mathbf{H}_{ji}^{\nabla}\mathbf{A}\mathbf{F}^{-1}\mathbf{v}_{i,\mathrm{o}}[t]
+\mathbf{H}_{ji}^{\triangle}\mathbf{A}\mathbf{F}^{-1}\mathbf{v}_{i,\mathrm{o}}[t-1]\big)$ as equivalent noise. Now, let 
\begin{align}
\mbox{cov}(\mathbf{z}_{j})&=N_{0}\mathbf{I}_{N+L}+\rho\alpha_{ji}\mathbf{H}_{ji}^{\nabla}\mathbf{A}\mathbf{F}^{-1}\mathbf{P}^{{\rm B}}_{i,\mathrm{o}}\mathbf{F}\mathbf{A}^{\mathrm{H}}\big(\mathbf{H}_{ji}^{\nabla}\big)^{\mathrm{H}}\notag
\\&\quad+\rho\alpha_{ji}\mathbf{H}_{ji}^{\triangle}\mathbf{A}\mathbf{F}^{-1}\mathbf{P}^{{\rm B}}_{i,\mathrm{o}}\mathbf{F}\mathbf{A}^{\mathrm{H}}\big(\mathbf{H}_{ji}^{\triangle}\big)^{\mathrm{H}}\label{Eq:P2_Without_Backhaul_NoiseCov}
\end{align}
be the corresponding noise-whitening matrix. Subsequently, we resort to the same approach adopted for the forward phase and write $\mbox{cov}^{-\frac{1}{2}}(\mathbf{z}_{j})\mathbf{H}_{s}^{\nabla}\mathbf{\Gamma}_{i}^{{\rm B}}
=\mathbf{U}_{i}^{{\rm B}}\mathbf{\Sigma}_{i}^{{\rm B}}\big(\mathbf{Q}_{i}^{{\rm B}}\big)^{\mathrm{H}}$, with $\mathbf{U}_{i}^{{\rm B}}\in\mathbb{C}^{(N+L)\times (N+L)}$ and $\mathbf{Q}_{i}^{{\rm B}}\in\mathbb{C}^{(N+L-\lvert\mathscr{N}_j\rvert)\times (N+L-\lvert\mathscr{N}_j\rvert)}$ unitary matrices, and 
$\mathbf{\Sigma}_{i}^{{\rm B}}=[\mbox{diag}(\sigma_{1,\mathrm{z}}^{{{\rm B}}}, \dots, \sigma_{N+L-\lvert\mathscr{N}_j\rvert,\mathrm{z}}^{{\rm B}}) \hspace{2mm} \mathbf{0}_{(N+L-\lvert\mathscr{N}_j\rvert)\times \lvert\mathscr{N}_j\rvert}]^\mathrm{T}$ where $\sigma_{i,\mathrm{z}}^{{\rm B}}$ is the $i$th singular value of $\mbox{cov}^{-\frac{1}{2}}(\mathbf{z}_{j})\mathbf{H}_{s}^{\nabla}\mathbf{\Gamma}_{i}^{{\rm B}}$. Accordingly, when no residual SI is present in the RX chain, the achievable rate over the link between the $i$th and the $j$th SNs in the backward phase is 
\begin{align}
&R^{{\rm B}}_{i,\mathrm{b}}\left(\rho\right)=\frac{1}{N+L}\notag
\\&\quad\times\sum_{n=1}^{N+L-\lvert\mathscr{N}_j\rvert}\log_2\left(1+\rho\alpha_{\mathrm{b}} \big[\mathbf{P}^{{\rm B}}_{i,\mathrm{b}}\big]_{nn}\left[\mathbf{\Sigma}_{i}^{{\rm B}}\right]^2_{nn}\right),
\label{Eq:P2_Without_Backhaul_Rate}
\end{align}
with $\big[\mathbf{P}^{{\rm B}}_{i,\mathrm{b}}\big]_{nn}=\big(\kappa_{5,i}-\big(\rho\alpha_{\mathrm{b}}\big[\mathbf{\Sigma}_{i}^{{\rm B}}\big]^2_{nn}\big)^{-1}\big)^{+},\,\forall\,n\in\{1,\cdots,N+L-\lvert\mathscr{N}_j\rvert\}$ WF solution. $\kappa_{5,i}$ is chosen such that $\mbox{Tr}\big(\mathbf{P}^{{\rm B}}_{i,\mathrm{b}}\big)=\left(N+L\right)P$. As before, since no residual SI is present in this case, $\rho^*=\frac{P_{\mathrm{th}}}{P}$. 

\subsubsection{Presence of Residual Self-interference}

We first plug $\alpha_{\mathrm{eq}}=\frac{N_0}{P_{\mathrm{th}}}$ into \eqref{Eq:P2_Split_Signal_Decoding_subtracting} and rewrite it as
\begin{equation}
\mathbf{y}_{j}^{{\rm SIC}}[t]=\sqrt{\rho\alpha_{\mathrm{b}}}\mathbf{H}_{s}^{\nabla}\mathbf{\Gamma}_{i}^{{\rm B}}\mathbf{C}_{i}^{{\rm B}}\mathbf{v}_{i,\mathrm{b}}[t]+\mathbf{z}_{j}[t],
\end{equation}
where
\begin{align*}
\mathbf{z}_{j}[t]&=\mathbf{w}_{j}[t]+\sqrt{\rho}\Big(\sqrt{\alpha_{ji}}\mathbf{H}^{\triangle}_{ji}\mathbf{A}\mathbf{F}^{-1}\mathbf{v}_{i,\mathrm{o}}[t-1]\notag
\\&\quad+\sqrt{\alpha_{ji}}\mathbf{H}_{ji}^{\nabla}\mathbf{A}\mathbf{F}^{-1}\mathbf{v}_{i,\mathrm{o}}[t]+\sqrt{\frac{N_0}{P_{\mathrm{th}}}}\mathbf{\Gamma}_{j}^{{\rm B}}\mathbf{C}_{j}^{{\rm B}}\mathbf{v}_{j,\mathrm{b}}[t]\bigg)
\end{align*}
is the equivalent noise. Subsequently, we let
\begin{align}
\mbox{cov}(\mathbf{z}_{j})&=N_{0}\mathbf{I}_{N+L}+\frac{N_0\rho\left(N+L\right)P}{P_{\mathrm{th}}(N+L-\left|\mathscr{N}_{i}\right|)}\mathbf{\Gamma}_{j}^{{\rm B}}\big(\mathbf{\Gamma}_{j}^{{\rm B}}\big)^{\mathrm{H}}\notag
\\&\quad+\rho\alpha_{ji}\Big(\mathbf{H}_{ji}^{\nabla}\mathbf{A}\mathbf{F}^{-1}\mathbf{P}^{{\rm B}}_{i,\mathrm{o}}\mathbf{F}\mathbf{A}^{\mathrm{H}}\big(\mathbf{H}_{ji}^{\nabla}\big)^{\mathrm{H}}\notag
\\&\qquad+\mathbf{H}_{ji}^{\triangle}\mathbf{A}\mathbf{F}^{-1}\mathbf{P}^{{\rm B}}_{i,\mathrm{o}}\mathbf{F}\mathbf{A}^{\mathrm{H}}\big(\mathbf{H}_{ji}^{\triangle}\big)^{\mathrm{H}}\Big)
\notag
\end{align}
be its covariance matrix, with $\mathbf{P}^{{\rm B}}_{j,\mathrm{b}}=\frac{\left(N+L\right)P}{N+L-\left|\mathscr{N}_{i}\right|}\mathbf{I}_{N+L-\left|\mathscr{N}_{i}\right|}$. Now, following the same procedures as in Sec.~\ref{Subsubsec: Down_With_Residual_Self-interference}, we write $\mbox{cov}^{-\frac{1}{2}}(\mathbf{z}_{j})\mathbf{H}_{s}^{\nabla}\mathbf{\Gamma}_{i}^{\rm B}=\mathbf{U}_{i}^{{\rm B}}\mathbf{\Sigma}_{i}^{{\rm B}}\big(\mathbf{Q}_{i}^{{\rm B}}\big)^{\mathrm{H}}$, with $\mathbf{U}_{i}^{{\rm B}}$,  $\mathbf{\Sigma}_{i}^{{\rm B}}$, and $\mathbf{Q}_{i}^{{\rm B}}$ defined analogously to the previous case. Finally, as before, we define $\mathbf{C}_{i}^{{\rm B}}=\mathbf{Q}_{i}^{{\rm B}}$ and let $\big(\mathbf{U}_{i}^{{\rm B}}\big)^{\mathrm{H}}\mbox{cov}^{-\frac{1}{2}}(\mathbf{z}_{j})$ be the decoding matrix for the $i$th signaling at the $j$th SN. Thus, when residual SI is present in the RX chain, the achievable rate over the link between the $i$th and the $j$th SN in the backward phase is 
\begin{align}
&R^{{\rm B}}_{i,\mathrm{b}}\left(\rho\right)=\frac{1}{N+L}\notag
\\&\quad\times\sum_{n=1}^{N+L-\left|\mathscr{N}_{j}\right|}\log_2\left(1+\frac{\rho\alpha_{\mathrm{b}} \left(N+L\right)P\left[\mathbf{\Sigma}_{i}^{{\rm B}}\right]^2_{nn}}{N+L-\left|\mathscr{N}_{j}\right|}\right).\label{Eq:P2_With_Backhaul_Rate}
\end{align}
As previously inferred, the values of $\rho^*$ for the backward phase are the same as for the forward phase Additionally, assessing the impact that $\rho$ may have on \eqref{Eq:P2_Without_Backhaul_Rate} and \eqref{Eq:P2_With_Backhaul_Rate}, when a WF strategy is adopted, is not straightforward. Hence, we can proceed as done for Propositions~\ref{Lem:Down_Lemma_Backhaul} and~\ref{Lem:Up_Lemma_OFDM} and obtain.
%
%
\begin{proposition}\label{Lem:Up_Lemma_Backhaul}
	The achievable rate for the transmission between the $i$th and the $j$th SN in the backward phase can be approximated as
	\begin{align}
	R_{i,\mathrm{b}}^{{\rm B}}\left(\rho\right)&\approx\begin{cases}
	\log_2\left(1+c_5\rho^{\phi_5}\right),\mbox{~when~} \rho\leq \frac{P_{\mathrm{th}}}{P},\\
	\log_2\left(1+c_6\rho^{\phi_6}\right),\\
	\hspace{2em}\mbox{when~} \frac{P_{\mathrm{th}}}{P}<\rho\leq\min\left\{1,\frac{P_{\mathrm{sat}}}{P}\right\},
	\end{cases}\label{Eq:Lem_3_first_case}
	\end{align}
	where
	\begin{align*}
	c_5&=\left(2^{ R_{i,\mathrm{b}}^{{\rm B}}(\frac{P_{\mathrm{th}}}{P})}-1\right)\left(\frac{P}{P_{\mathrm{th}}}\right)^{\phi_5}, 
	\\\phi_5&=\log_2\left(\frac{2^{ R_{i,\mathrm{b}}^{{\rm B}}(\frac{P_{\mathrm{th}}}{P})}-1}{2^{ R_{i,\mathrm{b}}^{{\rm B}}(\frac{P_{\mathrm{th}}}{2P})}-1}\right),
	\\c_6&=\left(2^{ R_{i,\mathrm{b}}^{{\rm B}}(\frac{P_{\mathrm{th}}+\epsilon}{P})}-1\right)\left(\frac{P}{P_{\mathrm{th}}+\epsilon}\right)^{\phi_6}, 
	\\\phi_6&=\log_{\frac{P_{\mathrm{th}}+\epsilon}{\min\{P,P_{\mathrm{sat}}\}}}\left(\frac{2^{ R_{i,\mathrm{b}}^{{\rm B}}(\frac{P_{\mathrm{th}}+\epsilon}{P})}-1}{2^{ R_{i,\mathrm{b}}^{{\rm B}}(\min\{1,\frac{P_{\mathrm{sat}}}{P}\})}-1}\right).
	\end{align*}
\end{proposition}
\begin{IEEEproof}
The result can be obtained as done for Proposition~\ref{Lem:Down_Lemma_Backhaul}.
\end{IEEEproof}
%
%
%
%
Now, applying Proposition~\ref{Lem:Up_Lemma_Backhaul} and following the same approach adopted for Corollary~\ref{Lem:Down_Remark_Backhaul}, we can state the following result.
%
%
\begin{corollary}\label{Lem:Up_Remark_Backhaul}
When $P_{\mathrm{th}}<P\leq P_{\mathrm{sat}}$, the interval of $\rho$ for which the achievable rate of the link between the two SNs is higher, if the proposed approach is adopted over the state-of-the-art solutions, can be approximated as $\bigg[\sqrt[\phi_5]{\frac{2^{R^{{\rm B}}_{i,\mathrm{b}}\left(1\right)}-1}{c_5}},\frac{P_{\mathrm{th}}}{P}\bigg]$.
\end{corollary}
%
%
\begin{IEEEproof}
The result can be obtained as done for Corollary~\ref{Lem:Down_Remark_Backhaul}.
\end{IEEEproof}

\vspace{-.6cm}
\subsection{Recycled Energy}\label{Subsec: Uplink harvesting}

Using the digital representation of the signals for the sake of compactness of the notation, and by some straightforward calculations, we can express the recycled energy per symbol at the $i$th SN as
\begin{align}
&E_{i}^{{\rm B}}\left(\rho\right)=\frac{\beta(1-\rho)}{N+L}\times\notag
\\&\left[\alpha_{ii}\mbox{Tr}\left(\left(\big(\mathbf{H}_{ii}^{\nabla}\big)^{\mathrm{H}}\mathbf{H}_{ii}^{\nabla}+\big(\mathbf{H}_{ii}^{\triangle}\big)^{\mathrm{H}}\mathbf{H}_{ii}^{\triangle}\right)\mathbf{A}\mathbf{F}^{-1}\mathbf{P}^{{\rm B}}_{i,\mathrm{o}}\mathbf{F}\mathbf{A}^{\mathrm{H}}\right)\right.\notag
\\&+\alpha_{ij}\mbox{Tr}\left(\left(\big(\mathbf{H}_{ij}^{\nabla}\big)^{\mathrm{H}}\mathbf{H}_{ij}^{\nabla}+\big(\mathbf{H}_{ij}^{\triangle}\big)^{\mathrm{H}}\mathbf{H}_{ij}^{\triangle}\right)\mathbf{A}\mathbf{F}^{-1}\mathbf{P}^{{\rm B}}_{j,\mathrm{o}}\mathbf{F}\mathbf{A}^{\mathrm{H}}\right)\notag
\\&+\alpha_{\mathrm{b}}\mbox{Tr}\Big(\left(\big(\mathbf{H}_{s}^{\nabla}\big)^{\mathrm{H}}\mathbf{H}_{s}^{\nabla}+\big(\mathbf{H}_{s}^{\triangle}\big)^{\mathrm{H}}\mathbf{H}_{s}^{\triangle}\right)\mathbf{\Gamma}_{j}^{{\rm B}}\mathbf{C}_{j}^{{\rm B}}\mathbf{P}^{{\rm B}}_{j,\mathrm{b}}\notag
\\&\times\big(\mathbf{C}_{j}^{{\rm B}}\big)^{\mathrm{H}}\big(\mathbf{\Gamma}_{j}^{{\rm B}}\big)^{\mathrm{H}}\Big)+\mbox{Tr}\Big(\Big(\alpha_{\mathrm{m}}\big(\mathbf{H}_{i}^{m\triangle}\big)^{\mathrm{H}}\mathbf{H}_{i}^{m\triangle}\notag
\\&+\left(\sqrt{\alpha_{\mathrm{c}}}\mathbf{I}_{N+L}+\sqrt{\alpha_{\mathrm{m}}}\mathbf{H}_{i}^{m\nabla}\right)^{\mathrm{H}}\left(\sqrt{\alpha_{\mathrm{c}}}\mathbf{I}_{N+L}+\sqrt{\alpha_{\mathrm{m}}}\mathbf{H}_{i}^{m\nabla}\right)\Big)\notag
\\&\times\mathbf{\Gamma}_{i}^{{\rm B}}\mathbf{C}_{i}^{{\rm B}}\mathbf{P}^{{\rm B}}_{i,\mathrm{b}}\big(\mathbf{C}_{i}^{{\rm B}}\big)^{\mathrm{H}}\big(\mathbf{\Gamma}_{i}^{{\rm B}}\big)^{\mathrm{H}}\Big)\Big].
\label{Eq:Up_Harvested_Energy_at_SN}
\end{align}
At this stage, we know from Sec.~\ref{sec:energy_forward} that a safe approximation of $E_{i}^{{\rm B}}\left(\rho\right)$ can be found by neglecting the received signaling and multi-path SI signals terms from \eqref{Eq:Up_Harvested_Energy_at_SN}. Thus, we can approximate the recycled energy per symbol at the $i$th SN as $E_{i}^{{\rm B}}\left(\rho\right)\approx \beta\alpha_{\mathrm{c}} \left(1-\rho\right)P$.

\vspace{-.2cm}
\section{Numerical Results}\label{Sec: Numerical results}

In this section, the performance of the proposed FD architecture is studied in the considered four-node setting. In particular, we consider a realistic indoor scenario in which we let the distance between the SNs and the ANs be $d_{\mathrm{sa}}=10$~m, and the distance between SNs be $d_{\mathrm{ss}} \in \{15, 20\}$~m. The transmitted signals are modulated at carrier frequency $f_\mathrm{c}=1800$~MHz, and experience a distance dependent path loss modeled according to~\cite{rpt:itu_indoor}. Furthermore, for the sake of consistency with the state of the art, we let $\alpha_{\mathrm{c}}=-10$~dB, $\alpha_{\mathrm{m}}=-35$~dB, $\beta=0.7$, $P_{\mathrm{sat}}=28$~dBm, and $P_{\mathrm{th}}=20$~dBm~\cite{conf:bharadia13, conf:knox12, art:Visser12}. As previously discussed, we are particularly interested in characterizing the behavior of the FD radio in critical conditions, i.e., when the state-of-the-art solutions fail to cancel all the SI in the RX chain. Thus, we let the TX power of the FD radio be $P\in [P_{\mathrm{th}},P_{\mathrm{sat}}]$, with $P_{\mathrm{o}}=P_{\mathrm{b}}=\frac{P}{2}$. Concerning the OFDMA transmission, we set $N=64$ and $L=16$ according to~\cite{rpt:3gpp.36.814}. We assume channels characterized by the exponentially decreasing PDP, such that $\displaystyle \xi_n^2=\exp\left\lbrace-\frac{nT_s}{\tau_h}\right\rbrace$ with $\tau_h$ the root mean square delay spread of the channel and $T_s$ the sampling period at each device. In particular, we restrict our study to channels with $l=L$ taps and the slowly decaying PDP, i.e., $\frac{T_s}{\tau_h}\in[2,3]$, for matters of space economy. 

Finally, we define the two metrics that may offer a straightforward way to identify the range of values of $\rho$ for which the proposed FD architecture outperforms the state-of-the-art solution \cite{conf:bharadia13}. In particular, we let $\eta_R=\frac{R^{\cdot}_{i,\cdot}\left(\rho\right)}{R^{\cdot}_{i,\cdot}\left(1\right)}$, and  $\eta_E=\frac{E_{i}^{\cdot}\left(\rho\right)}{E_{tx}}$, with $E_{tx}$ energy per transmitted symbol by the FD device. These metrics can be directly interpreted as follows. If the proposed FD architecture achieves a larger spectral efficiency than the state of the art, then $\eta_R>1$. Similarly, if a portion of the energy of the transmitted signal is actually recycled by the proposed architecture, then $\eta_E>0$.

\vspace{-.4cm}
\subsection{Forward Phase}

\begin{figure}[!h]
\vspace{-.2cm}
	\centering
    \includegraphics[width=\columnwidth]{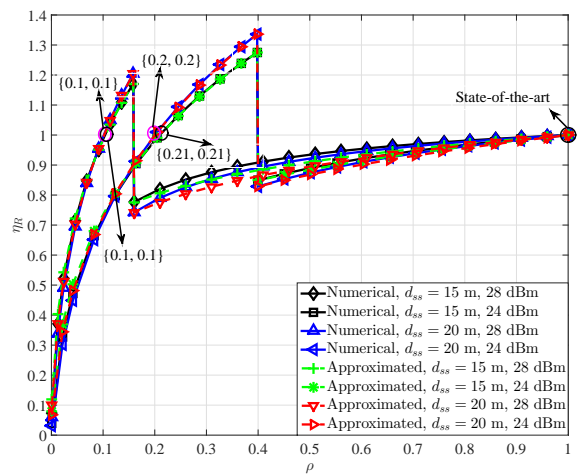}
\vspace{-.7cm}	
	\caption{[Forward] Numerical and approximated $\eta_R$ for the signaling as $\rho$ varies.}
    \label{fig:forward_WB}
\end{figure}

\begin{figure}[!]
\centering
\includegraphics[width=\columnwidth]{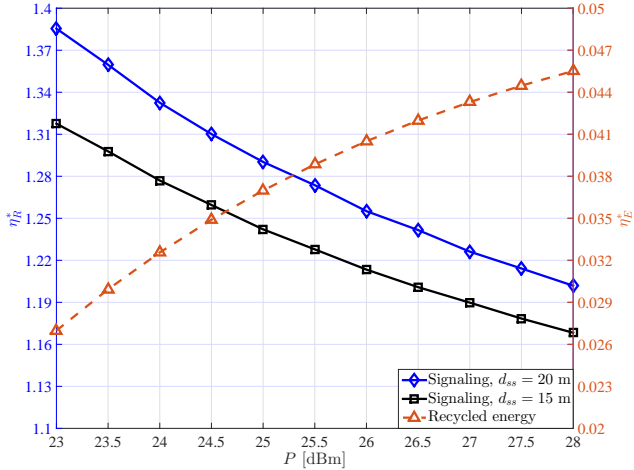}
\vspace{-.7cm}	
\caption{[Forward] $\eta^*_R$ and $\eta^*_E$ as the transmit power $P$ varies.}
\label{fig:forward_etas}
\end{figure}

We first focus on the forward phase. We recall that no transmission is performed by the ANs in this phase. For the first study, we consider only two values for $P$ within the aforementioned range, i.e. $P=24$~dBm and $P=28$~dBm, for  ease of representation. Subsequently, we let $\rho\in[0,1]$ and compute $\eta_R$, both analytically, e.g., by means of the derived approximations, and numerically, in Fig.~\ref{fig:forward_WB}. 
We start by observing that a range of values of $\rho$ for which $\eta_R\geq 1$ can be found in each of the tested configurations. In this regard, we note that the couples of numbers (in the form $\{\cdot, \cdot\}$) depicted in the figure denote the numerical and approximated lower bounds of the aforementioned range of values of $\rho$, i.e., $\eta_R\geq 1, \hspace{2mm} \forall \rho \in [\{\cdot, \cdot\}, \rho^*]$. At a first glance, we observe that our approximations do not provide the same accuracy for each value of $\rho$. However, they perfectly match the numerical values for $\rho \in [\{\cdot, \cdot\}, \rho^*]$, which is the most relevant region in terms of system design insights. This is a remarkable result since it allows the system designer to design methods to set suitable values of $\rho$ on-the-fly, depending on the performance target, without the need for long offline simulations. Quantitatively, the TX power of the FD can be significantly increased as compared to the state of the art. In fact, a performance increase up to $35\%$ and $21\%$ is achieved for $P=24$~dBm and $P=28$~dBm, respectively. In practice, the larger $P$, the smaller $\rho$, i.e., the lower the power of the IC of the received signal and the resulting equivalent SNR. In this sense, the benefits brought by the proposed architecture are evident, even though they decrease as the TX power increases. An in-depth analysis of the theoretical limits of the performance gain is the matter of our future research. Finally, we note that the impact of $P$ on $\eta_R$ is larger than the impact of $d_{\mathrm{ss}}$. Intuitively, this can be explained as follows. According to the adopted path loss model~\cite{rpt:itu_indoor}, the received signal at the SN experiences a reduction of $~6$~dB per octave. Conversely, the SI suffered by the device increases at a much higher pace if the TX power is augmented, in turn reducing $\eta_R$. Given the extent of the impact of $P$ on the performance of the FD radio, we now consider $P\in[23,28]$~dBm and compute the optimal $\rho$, i.e., $\rho^*$, for each of the considered values. Subsequently, we compute both $\eta_R$ and $\eta_E$ when $\rho^*$ is adopted, and depict them in Fig.~\ref{fig:forward_etas} as $\eta^*_R$ and $\eta^*_E$ for consistency with the notation.
We start by noting that $\eta^*_R>1$ for each of considered configurations. Moreover, as previously inferred, we see that $\eta^*_R$ increases as $P$ decreases. In particular, we observe that a TX power increase of $5$~dB induces a reduction of $15\%$ and $18\%$ for $d_{\mathrm{ss}}=15$~m and $d_{\mathrm{ss}}=20$~m, respectively. Similarly to what has been observed in Fig.~\ref{fig:forward_WB} for $\eta_R$, $\eta^*_R$ also increases with $d_{\mathrm{ss}}$. In practice, the lower the power of the received signal at the SNs, the higher the performance gain the proposed FD architecture can deliver. This interesting result could have been expected. In fact, the impact of an effective SIC on the achievable rate of the incoming signal is larger when the latter experiences a severe path loss during its propagation (i.e., the power of the received signal by the FD radio is lower). We now switch our focus to $\eta^*_E$. In Fig.~\ref{fig:forward_etas}, its value ranges between $2.7\%$ and $4.7\%$ for the considered values of $P$. In other words, up to around $50\%$ of the leaked energy can be recycled as compared to the state of the art, while guaranteeing better performance in terms of throughput. Quantitatively, $\eta^*_E$ is not influenced by a change of $d_{\mathrm{ss}}$, since it uniquely depends on the energy of the signal leakage at the circulator. Additionally, we note that the aforementioned values should actually be seen as lower bounds for $\eta_E$, since they are obtained when $\rho^*$ is adopted. In fact, if any other value of $\rho$ within the range $[\{\cdot, \cdot\}, \rho^*]$ was adopted, then the resulting $\eta_E$ would be larger (in turn reducing $\eta_R$). This further confirms the potential of the proposed architecture, whose benefits in the forward phase can be summarized as: 1) higher achievable rate for the transmission towards the AN (thanks to the higher TX power), 2) higher achievable rate for the incoming signaling, and 3) lower power consumption.
\vspace{-.4cm}

\subsection{Backward Phase} 

\begin{figure}[!h]
\vspace{-.2cm}
\centering
\includegraphics[width=\columnwidth]{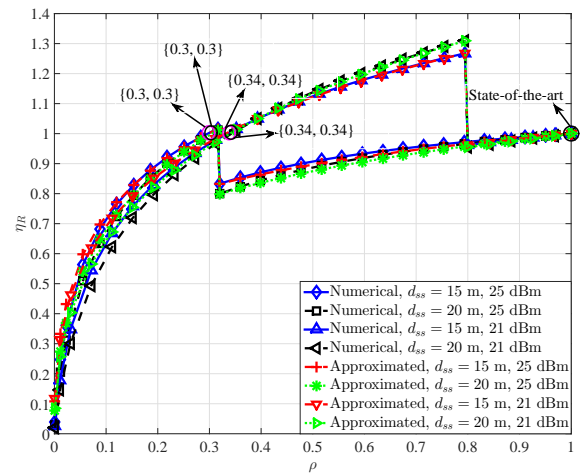}
\vspace{-.7cm}	
\caption{[Backward] Numerical and approximated $\eta_R$ for the signaling as $\rho$ varies.}
\label{fig:bacward_signaling}
\end{figure}

Differently from the forward phase, both SNs and ANs transmit during the backward phase. In this regard, we note that no OFDMA signal is transmitted by the SNs in this phase. Thus, we will lower the values of $P$ considered in the previous section by $3$~dB, to make sure that the same TX power is invested for the signaling in both phases, and obtain a fair comparison. As before, we start our study by focusing on the achievable rate for the signaling between the two SNs and compute $\eta_R$ as $\rho$ varies in Fig.~\ref{fig:bacward_signaling}.
Qualitatively, both the values of $\eta_R$ for the signaling and the accuracy of our approximations in the backward phase are extremely similar to what we observed for the forward phase. Accordingly, the impact of $P$ on $\eta_R$ is larger than the impact of $d_{\mathrm{ss}}$, and the larger $P$, the lower the resulting $\eta_R$. However, we note that the performance gain brought by the proposed architecture over the state of the art in this case is quantitatively lower. This result can be understood by comparing \eqref{Eq:Down_Without_Backhaul_EqivNoise_Cov} and \eqref{Eq:P2_Without_Backhaul_NoiseCov}. Therein we see that side knowledge about the covariance matrix of the interference generated to the signaling by the OFDM transmission can be safely considered available during the forward phase. The same is not true for the backward phase, where such information is hardly available, unless non-linear and/or complex iterative decoding procedures are adopted at the SNs. Thus, the achievable rate for the signaling is lowered by the presence of the interference caused by the OFDM signals transmitted by the ANs. We will further discuss this aspect in the last study of the section.

\begin{figure}[!h]
\vspace{-.2cm}
\centering
\includegraphics[width=\columnwidth]{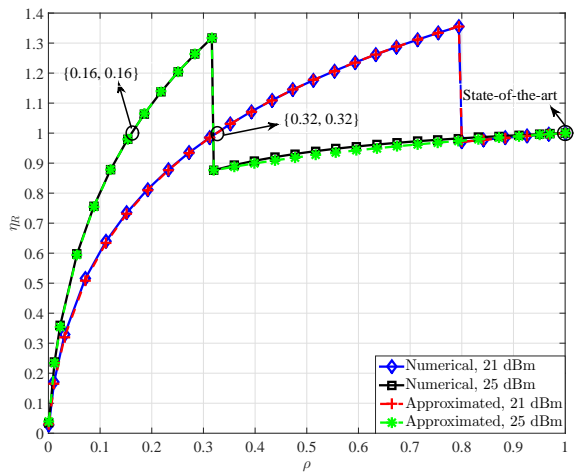}
\vspace{-.7cm}	
\caption{[Backward] Numerical and approximated $\eta_R$ for the backward signals as $\rho$ varies.}
\label{fig:backward_UL}
\end{figure}

\begin{figure}[!h]
\vspace{-.2cm}
\centering
\includegraphics[width=\columnwidth]{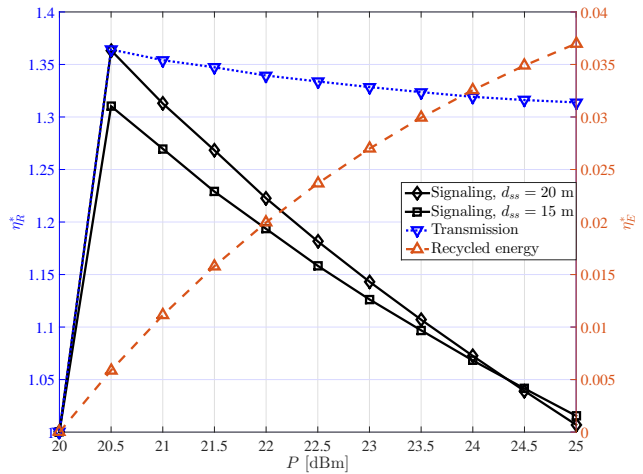}
\vspace{-.7cm}	
\caption{[Backward] $\eta^*_R$ and $\eta^*_E$ as the transmit power $P$ varies.}
\label{fig:backward_etas}
\end{figure}

As a matter of fact, the signals transmitted by the ANs are not only a source of interference for the signaling, but convey information for the SNs as well. Hence, in the next study we will focus on their achievable rates and compute $\eta_R$ w.r.t.~these transmissions, as $\rho$ varies, and depict it in Fig.~\ref{fig:backward_UL}.
Interestingly, the qualitative behavior of $\eta_R$ for the backward signals is very similar to the behavior for the signaling in both forward and backward phases. Remarkably, the accuracy of our approximations is higher in this case, substantiating even more their importance in terms of system design insights. Quantitatively, the obtained results are significantly better than the performance increase achieved for the signaling in the backward phase. This could have been expected since no interference is affecting the OFDMA decoding at the SNs, provided that the signaling satisfy the constraints in \eqref{Eq:P2_Backhaul_Constraint_Dumb}. The absence of interference also implies that the impact of a TX power change on $\eta_R$ is less evident in this case than in any of the previous tests. In particular, we note that the performance gain over the state of the art is rather remarkable, i.e., between $31\%$ and $37\%$ for both considered values of $P$. Now, as for the forward phase we conclude the analysis of the backward phase by letting $P\in[20,25]$~dBm and compute  $\rho^*$ for each of the considered values and transmissions (i.e., signaling and backward signals). Analogously, $\eta^*_R$ and $\eta^*_E$ are computed and depicted in Fig.~\ref{fig:backward_etas}.
We first focus on the signaling. As for the forward phase, $\eta^*_R$ decreases as $P$ increases. This is due to the fact that $\rho^*$ decreases as $P$ increases, in turn reducing the effective SNR of the IC of the received signal, and thus the achievable rate for the signaling. Additionally, as noted for Fig.~\ref{fig:bacward_signaling}, $\eta^*_R\approx 1$ for $P=25$~dBm, due to the aforementioned impact of the interference generated by the OFDM signals transmitted by the ANs. We note that this issue is more evident as the distance between the SNs increases, as intuitively it should be. Switching our focus to the backward signals, we note a very consistent behavior of $\eta^*_R$ as $P$ increases, as expected. The decreasing trend of $\eta^*_R$ is still present, but with a much lower pace. In a way, this confirms the potential of the proposed architecture in terms of the performance gain w.r.t.~the state of the art and highlights its merit in terms of overall throughput enhancement for the considered system, regardless of the considered phase of the transmission. Finally, concerning the amount of energy recycled in this phase, we note that values up to $\eta^*_E=3.7\%$ are obtained in this phase. In practice, this implies that almost $40\%$ of the leaked energy can be recycled as compared to the state of the art, while guaranteeing better performance in terms of throughput. The same observations made for Fig.~\ref{fig:forward_etas} in the study of the forward case hold for the backward phase as well, hence they will not be repeated.

\section{Conclusions}\label{Sec: Conclusions}

In this work, we have introduced a novel energy-recycling FD radio architecture that can provide spectral efficiency and energy consumption improvements over the state of the art. The performance gain is achieved thanks to the introduction of a 3-port element between the circulator and the RX chain including a power divider and an RF energy harvester. The impact of this element is two-fold. First, it allows for an arbitrary attenuation of the incoming signal, in turn increasing the effectiveness of the state-of-the-art SIC strategies subsequently adopted in the RX chain. Second, it recycles a non-negligible portion of the energy leaked through the non-ideal circulator. We have characterized the performance of this architecture in a hybrid FD/HD four-node network, framed according to practically relevant considerations, in which 2 nodes operate in FD and 2 nodes in HD. More specifically, we provide analytical approximations of: 1) the achievable rates for the transmissions performed by the FD and HD radios as the direction of the communication with the HD radios changes, e.g., the forward and backward phases, and 2) the amount of energy recycled by the FD radio in these phases. The accuracy of these derivations has been substantiated by our numerical findings, by which the gains that the proposed architecture can yield over its state-of-the-art alternatives have been illustrated. The next steps of our research are the design of a variant of the proposed FD architecture in the presence of multiple antennas, and an in-depth analysis of the theoretical limits of the performance gain brought by its adoption.

\bibliographystyle{IEEEtran}
\bibliography{ref}

\end{document}